\documentclass[12pt]{article}
\pdfoutput=1
\usepackage{amsmath,amssymb}
\usepackage{graphicx}
\usepackage{cite,fancyhdr}

\usepackage[affil-it]{authblk}
\usepackage[left=2cm,right=2cm,top=3cm,bottom=3cm,a4paper]{geometry}

\newcommand{\Slash}[1]{{\ooalign{\hfil#1\hfil\crcr\raise.167ex\hbox{/}}}}

\newcommand{\beq}{\begin{equation}}  \newcommand{\eeq}{\end{equation}}
\newcommand{\bef}{\begin{figure}}  \newcommand{\eef}{\end{figure}}
\newcommand{\bec}{\begin{center}}  \newcommand{\eec}{\end{center}}
\newcommand{\non}{\nonumber}  
\newcommand{\laq}[1]{\label{#1}}

\newcommand{\Eqs}[1]{Eqs.(\ref{#1})}
\newcommand{\eq}[1]{(\ref{#1})}

\newcommand{\ab}[1]{\left|{#1}\right|}
\newcommand{\vev}[1]{ \left\langle {#1} \right\rangle }

\newcommand{\SU}[1]{{ SU{#1} } }

\def\({\left(}
\def\){\right)}

\def\O{\mathcal{O}}
\def\U{\mathop{U}}

\newcommand{\OR}{~{\rm or}~}
\newcommand{\AND}{~{\rm and}~}

\newcommand{\GEV}{ {\rm \, GeV} }
\newcommand{\TEV}{ {\rm \, TeV} }
\def\o{\over}
\def\a{\alpha}
\def\b{\beta}

\def\d{\delta}
\def\e{\epsilon}
\def\f{\phi}
\def\g{\gamma}

\def\k{\kappa}

\def\m{\mu}

\def\t{\tau}

\def\F{\Phi}

\def\tl{\tilde}
\def\*{\dagger}

\begin{document}

\allowdisplaybreaks
\setcounter{footnote}{0}
\setcounter{figure}{0}
\setcounter{table}{0}

\title{\bf \large 
Bino-wino coannihilation as  a prediction in the $E_7$ unification of families}

\author[1,2]{{\normalsize  Tsutomu T. Yanagida}}
\author[3]{{\normalsize Wen Yin}}
\author[4]{{\normalsize Norimi Yokozaki}}

\affil[1]{\small 
T.D.Lee Institute and School of Physics and Astronomy, 

Shanghai Jiao Tong University, Shanghai 200240, China}

\affil[2]{\small Kavli IPMU (WPI), UTIAS, The University of Tokyo, 

 Kashiwa, Chiba 277-8583, Japan}

\affil[3]{\small 
Department of Physics, KAIST, Daejeon 34141, Korea}

\affil[4]{\small 
Department of Physics, Tohoku University,  

Sendai, Miyagi 980-8578, Japan}

\date{}

\maketitle

\thispagestyle{fancy}
\rhead{TU-1089  }
\cfoot{\thepage}
\renewcommand{\headrulewidth}{0pt}

\begin{abstract}
\noindent
We study the phenomenological consequences of the supersymmetric (SUSY) $E_7/\SU(5)\times \U(1)^3$ non-linear sigma model coupled to supergravity, where the three generations of quark and lepton chiral multiplets appear as (pseudo) Nambu Goldstone (NG) multiplets, that is, the origin of the three families is explained. 
To break SUSY, we introduce a SUSY breaking field charged under some symmetry avoiding the Polonyi problem. 
The gaugino mass spectrum is almost uniquely determined 
when one requires the electroweak vacuum to be (meta)stable:
it would be a miracle that the mass difference between the bino and wino turns out to be within $ \O(1)\%$ at the low energy. Thus, a bino-wino coannihilation is naturally predicted, which can explain the correct relic abundance of dark matter.
Moreover, we find that the bottom-tau Yukawa couplings and the gauge couplings are unified up to $\O(1)\%$ in most of 
the viable region. This scenario can be fully tested at the LHC and future collider experiments since the gauginos and some of the pseudo-NG bosons are light. 
An axion-like multiplet, which can be identified with the QCD axion, is also predicted.
\end{abstract}

\clearpage
\section{Introduction}

The presence of three families of quarks and leptons is one of the fundamental questions in nature: why do we have three families? 
It was shown that the supersymmetric (SUSY) non-linear sigma (NLS) model of $G/H=E_7/SU(5) \times U(1)^3$~\cite{Kugo:1983ai, Yanagida:1985jc} accommodates three families of quarks and leptons as Nambu-Goldstone (NG) multiplets of the symmetry breaking. 
In fact, the approach of NLS models based on exceptional groups predicts the maximal number of families, since the exceptional group is limited up to $E_8$.  It was shown that the number of the families is limited to be three even if we take the biggest exceptional group $E_8$~\cite{Irie:1983cd}. The scenario is fascinating because it answers the fundamental question why nature has three families of quarks and leptons.

The unbroken $SU(5)$ is identified with the gauge group of the grand unified theory (GUT). The symmetry $E_7$ is explicitly broken by the gauge couplings and the Higgs Yukawa couplings to the quarks and leptons. Once SUSY is broken, the NG bosons (NGBs) get soft SUSY breaking masses proportional to these couplings at the loop-level and become pseudo-NGBs (pNGBs). 
On the other hand, Higgs doublets, $H_u$ and $H_d$, are not (p)NGBs, and hence, their soft SUSY breaking masses are not suppressed.  
This scenario together with pure gravity mediation (PGM)~\cite{Ibe:2006de,Ibe:2011aa} or minimal split SUSY~\cite{ArkaniHamed:2012gw} 
was studied in Ref.~\cite{Yanagida:2016kag}. It was shown that the Higgs mediation~\cite{Yamaguchi:2016oqz} and anomaly mediation~\cite{Giudice:1998xp,Randall:1998uk} arising from this framework lead to the so-called Higgs-anomaly mediation~\cite{ Yin:2016shg,Yanagida:2016kag}. In this scenario, the muon anomalous magnetic moment ($g-2$) anomaly, as well as the bottom-tau or top-bottom-tau Yukawa coupling unification, can be explained. Light sleptons and squarks are predicted in a way that the flavor changing neutral currents are suppressed~\cite{Yanagida:2018eho}.

In this paper, we first revisit a general NLS model on a compact K\"{a}hler manifold coupled to supergravity, and point out there generally exist important SUSY breaking effects. 
According to Refs.~\cite{Komargodski:2010rb, Kugo:2010fs}, a chiral multiplet, $S$, must exist to preserve the $G$-invariance. The model has a shift symmetry
\beq \laq{1} S\to S+i\a\eeq
where $\a$ is a real constant.
We point out that once SUSY is broken by an $F$-term of a SUSY breaking field $Z$, $S$ also acquires an $F$-term 
\beq
F_S=m_{3/2}f_s
\eeq
on a basis where the NGBs and $S$ are canonically normalized. 
Here, $f_s$ is a dimension one constant representing the typical size of the higher dimensional coupling of $S$ and $m_{3/2}$ is the gravitino mass; we have assumed that $Z$ is charged under some symmetry and a vacuum expectation value of the scalar component of $Z$ is vanishing.
Since $S$ must have couplings to the NG multiplets suppressed by  $1/f_s$ for the $G$-invariance, 
the $F$-term plays an important role in the SUSY breaking mediation via $S$ multiplet ($S$-mediation) in various models. 

In particular, in the $E_7/\SU(5) \times \U(1)^3$ NLS model with the SUSY breaking field, $Z$, charged under some symmetry, the Polonyi problem is absent,\footnote{This kind of SUSY breaking field is assumed in PGM~\cite{Ibe:2006de},  anomaly mediation~\cite{Giudice:1998xp}, and so on.}
and interesting phenomena are predicted taking into account $S$-mediation. 
Much below the scale $f_s$, this model has the particle contents of the minimal supersymmetric standard model (MSSM) plus the weakly-coupled $S$ multiplet. 
The gaugino masses are suppressed at the tree-level and they dominantly arise from anomaly mediation and $S$-mediation at the one-loop level.
The former contribution is known to be ultra-violet (UV) insensitive and the latter contribution is shown to be almost UV insensitive with given NG multiplets.
It turns out that the ratios of the gaugino masses only depend on the Higgs couplings to $S$ in the K\"{a}hler potential. 
Surprisingly, in most of the region compatible with vacuum (meta)stability of the electroweak minimum 
and the Higgs boson mass, 
the mass difference between the bino and wino is predicted to be within $\O(1)\%$. 
Therefore, the correct relic abundance of dark matter is naturally explained by the bino-wino coannihilation.
The range of the gravitino mass consistent with the dark matter abundance is found to be
$$40\TEV\lesssim m_{3/2}\lesssim 150\TEV.$$  
Furthermore,  in most of the mass range the bottom-tau Yukawa coupling unification and gauge coupling unification occur 
at $\O(1)\%$ level.
This scenario can be fully tested at the LHC and future collider experiments by searching for colored SUSY particles.  We also point out that the boson of $\Im{[S]}$ is consistent with the QCD axion solving the strong CP problem.
The moduli and gravitino problems are also discussed.

This paper is organized as follows. In section 2, we  revisit the NLS model in the framework of supergravity and discuss the $F$-term of the $S$ multiplet. 
In section 3, we consider the NLS model of $G/H=E_7/\SU(5) \times \U(1)^3$ and show the mass spectra of the MSSM particles. 
In section 4, the phenomenological consequences of the scenario are shown including the prediction of the bino-wino coannihilation and the coupling unifications.
The possible identification of $\Im[S]$ with the QCD axion is discussed in section 5.  
The last section is devoted to conclusions and discussions.

\section{Non-linear sigma (NLS) model in Supergravity}
\subsection{Review on NLS model}
Let us consider a NLS model defined in a compact K\"{a}hler manifold $G/H$ in supergravity. 
The K{\" a}hler potential for the NG multiplets is constructed from a real function of dimension two transforming under $G$ as
\begin{eqnarray}
\laq{ginv1}
\mathcal{K}(\phi_i, \phi_j^\dag) \to \mathcal{K}(\phi_i, \phi_j^\dag) + f_H(\phi_i) + f_H(\phi_i)^\dag,
\end{eqnarray}
where $f_H$ is a holomorphic function of NG multiplets $\phi_i$, and $\mathcal{K}$ is invariant under the transformations of the unbroken symmetry, $H$. The function $\mathcal{K}$ can be written as
\beq
\mathcal{K}(\phi_i, \phi_j^\dag)= \phi_i^\dag \phi_i+\cdots.
\eeq
where $\cdots$ denote the higher order terms of $\f_i, \f_j^\*.$
The real function $\mathcal{K}$ itself is not $G$ invariant and the shift, $f_H(\phi_i) + h.c.$, changes the Lagrangian in the framework of supergravity; therefore, there must exist a chiral superfield, $S$, canceling the shift~\cite{Komargodski:2010rb, Kugo:2010fs}. 
We can construct the $G$-invariant K{\" a}hler potential in a general form
\begin{eqnarray}
\laq{K}
K(\phi_i,\phi_j^\dag,S,S^\dag) =
f_{s0}^2 F\(X\), \label{eq:nlsm}
\end{eqnarray}
with 
\beq X\equiv\frac{\mathcal{K}(\phi_i, \phi_j^\dag)}{f_{s0}^2}  +  \frac{S}{f_{s0}} +\frac{S^\dag}{f_{s0}}\eeq
being a $G$-invariant. Here, $f_{s0}$ is a dimension one constant which characterizes the typical scale of the higher dimensional couplings of $S, \f_i, \AND \f_i^\*$. 
In general, the scale of higher dimensional couplings among $\phi_i, \phi_i^\*$ can differ from $f_{s0}$, but this does not change our conclusion (see Appendix\,\ref{ape1}). 
The shift symmetry of $S$ appears as 
\begin{eqnarray}
S \to S -\frac{1}{f_{s0}} f_H(\phi_i),
\end{eqnarray}
under $G$ transformation.  

Since the NLS model of Eq.~(\ref{eq:nlsm}) is consistently coupled to supergravity, the fields, $\f_i$, are massless from the NG theorem even when SUSY is broken. This can be alternatively understood from the fact that the scalar potential is also a function respecting the $G$-invariance~\cite{Goto:1990me}:
\beq
V=V(X), \label{eq:scalar_potential}
\eeq
(The direct derivation of the supergravity potential of a NLS model is given in Appendix\,\ref{ape1}.) If $V$ is stabilized at $X=\vev{X}$ with $\vev{}$ being the vacuum expectation value (VEV), we get
\beq \left. \frac{\partial{V}}{\partial X}\right|_{X=\vev{X}}=0.\eeq
Then, one obtains 
\beq
\left.\frac{\partial V}{\partial{\cal K}}\right|_{{X}=\vev{X}}=\frac{1}{ f^2_{s0}}\left.\frac{\partial V}{{\partial  X}}\right|_{{X}=\vev{X}}=0,
\eeq
which means that the coefficients of $|\f_i|^2 $, i.e. the mass terms, vanish.

\subsection{$F$-term of $S$}

Now let us assume that\footnote{Our main conclusion, the quasi-degenerate bino and wino, does not change even if $f_{s0}\sim M_P$}
\beq
f_{s0}\ll M_P.
\eeq
In this case, we have to introduce a SUSY breaking field $Z$ with \beq \vev{F_Z} \simeq \sqrt{3} m_{3/2} M_P.\eeq
Here, we assume $Z$ is charged under some symmetry and the VEV of the scalar component vanishes.
Although we can include direct couplings of $\f_i, 
\f_i^\*$ and $S$ to $Z$, the masslessness of $\phi_i$ is guaranteed as long as the couplings do not violate the $G$-invariance. 
The multiplet $S$ acquires an $F$-term once SUSY is broken. 
In particular, we will show that,
because of the $G$-invariance, the size of $F$-term is fixed to be a specific value when kinetic terms for $S$ and $\phi_i$ are canonically normalized.

Let us canonically normalize the kinetic terms by field redefinitions. First, we redefine $S$ such that it has the vanishing VEV
 \beq \vev{S}=0.\eeq
Then, the K\"{a}hler potential can be expanded as \cite{Siegel:1978mj, Kugo:1982cu, Kugo:1983mv}
\begin{eqnarray}
\laq{kine}
&& -3M_P^2 \F_c^\* \F_c  \exp{\(-\frac{K}{3M_P^2}\)} \nonumber \\
&\ni& \F_c^\* \F_c \(c_1 \(\f^\*_i\f_i+{f}_{s0} S+f_{s0} S^\* \)+\frac{c_2}{f_{s0}^2} \({\f^\*_i\f_i}+f_{s0} S+f_{s0}  S^\*\)^2 \) 
+\cdots,
\end{eqnarray}
where $c_1$ and $c_2$ are functions of $|Z|^2$ and we introduced the compensator multiplet of $\F_c=1-\theta^2 m_{3/2}.$ 
The combination in the brackets is required from the $G$-invariance; 
$\cdots$ denotes higher order terms of $(\f_i^\* \f_i +f_{s0}  S+f_{s0}  S^\*)$, terms only with $|Z|^2$, and the constant term $-3M_P^2 \F_c^\* \F_c$.  
By the field redefinition of $\f_i$ and $S$, the kinetic terms can be normalized as
\beq
\laq{kine2}
\eq{kine}\to  \f^\*_i\f_i+  S^\* S  + f_s \F_c ^\* S+f_s  S^\*\F_c  + 2\frac{\Re[ S \F_c^{-1}]}{f_s}\f^\*_i \f_i 
\eeq
where we omit terms irrelevant to the following discussion; 
we have defined $f_s \equiv (c_1/\sqrt{2c_2}) f_{s0}$. 
We emphasize that the coefficients of the linear terms of $ S$ and $ S^\*$ are fixed by the $G$-invariance under which $f_s  S  +f_s  S^\* +\phi_i^\* \phi_i +\cdots$ is invariant.
From the equation of motion for $F_S$, one immediately gets 
\beq
\laq{F}
F_{S}=f_s m_{3/2}.
\eeq
Since $S$ couples to the $\f_i$ at $1/f_s$, the SUSY breaking mediation from $S$-multiplet is sizable even though $F_S \ll M_P m_{3/2}.$ 
Consequently, the $S$-mediation plays an important role in a realistic model where $G$-invariance is explicitly broken by the gauge and Higgs Yukawa couplings to the NGBs: $A$-terms are generated at the tree-level, and gauginos and the masses of NGBs are generated at loop level.

\section{Mediation mechanisms predicted by $E_7/\SU(5)\times \U(1)^3$ NLS model}
In this section, we study the SUSY particle mass spectra of $G/H=E_7/SU(5)\times U(1)^3$ NLS model by taking into account the $F$-term of the $S.$
The $E_7/\SU(5) \times \U(1)^3$ model explains why there are three families of the leptons and quarks. 
There are $133-24-3=106$ NG modes:
\begin{eqnarray}
\phi_a^I: {\bf \bar 5}, \
\phi_I^{ab}: {\bf 10}, \
\phi_I : 1,\ \phi^a: {\bf 5},
\end{eqnarray}
where $a,b=1 \dots 5$ and $I=1 \dots 3$. Here, $\phi_a^I$ and $\phi_I^{ab}$ are identified with chiral multiplets of leptons and quarks in the MSSM, and $\phi_I$ are the right-handed neutrino multiplets. One finds that the three families of the leptons, quarks and right-handed neutrinos appear as NG multiplets. The charges for $U(1)^3$ are given in \cite{Sato:1997hv}. Strictly speaking, one needs an additional matter multiplet (non-NGB), ${\bf \bar 5'}$, to cancel the non-linear sigma model anomaly~\cite{Yanagida:1985jc}.
To be realistic, we assume that $E_7$ is explicitly broken by gauge and Higgs Yukawa couplings to the NGBs. Then, we can identify $\SU(5)$ in $H$ with the GUT gauge group: $\f^\* \f\to \f^\*e^{2g_5 V_5}\f$, where $V_5$ is 
the vector supermultiplet of $\SU(5),$ and $g_5$ is the coupling constant.

Since we can write down a Dirac mass term between $\f^a$ and ${\bf \bar 5'}$, they are decoupled in the low energy theory.
Moreover, Majorana mass terms for $\phi_I$ are allowed, which make the right-handed neutrinos heavy.\footnote{These explicit breaking terms of the $E_7$ can be obtained by Yukawa couplings to gauge singlet Higgs multiplets with non-vanishing VEVs~\cite{Sato:1997hv}. } By integrating out $\phi_I$, tiny neutrino masses and neutrino oscillations are explained by the seesaw mechanism~\cite{Yanagida:1979as, GellMann:1980vs, Glashow:1979nm} (see also Ref.~\cite{Minkowski:1977sc}).
The GUT gauge group is broken down to the SM gauge group $\SU(3)_c\times \SU(2)_L\times \U(1)_Y$ around $10^{16}\GEV.$
At this scale, there may be various multiplets relevant to the GUT breaking. However,  
in what follows, we will focus on the effective theory after the decouplings of the heavy degrees of freedom, which are much heavier than the gravitino mass scale. 
The effect of integrating out the heavy particles will be shown to be irrelevant to our main results.

The Yukawa interactions involving the Higgs doublets, $H_u$ and $H_d$, which are matter (non-NG) multiplets of $E_7$, are given as follows:
\beq
\laq{super}
W\ni
 y_u H_u Q u+y_d H_d Q d+y_e H_d L e,
\eeq
where we have omitted the flavor indices, and $\f_{I}^{ab}= \{Q_I, u_I, e_I \}, ~\f^{I}_{a}= \{d_I , L_I \}.$ 
Here we have assumed a discrete $R$-symmetry, $Z_{4R}$, where the NG multiplets should carry zero charge, and the Higgs doublets carry two.  
As an accidental symmetry, the ordinary $R$-parity appears.

The $E_7$-invariant K\"{a}hler potential at the zero limit of the gauge couplings relevant to our discussion is
\beq
\laq{kahler}
K \ni  x+\frac{1}{2f_s^2}x^2 +\hat{c}_u|H_u|^2 +\hat{c}_d |H_d|^2 +\hat{c}_\mu H_u H_d +h.c.
\eeq
where we define $x=\f_i^\* \f_i+f_s  S + f_s S^\*$ with $\f_i$ being the NG multiplets in the MSSM, 
we have not included the terms with fields of higher power. 
The coefficients, $\hat{c}_{u,d}=1+c_{u,d} x+ c_{u,d}^{(2)} x^2/2+\O(x^3)$ and $\hat{c}_\mu=c_\m+c_\m^{(1)} x+c_\m^{(2)} x^2/2 +\O(x^3)$, are functions of $x$. 

Now let us introduce a SUSY breaking field $Z=\theta^2 \sqrt{3} M_P m_{3/2}$ charged under some symmetry so that the Polonyi problem is absent.
 Then, $ S \AND  S^\*$ in $x$ acquire the $F$-term \eq{F}. 
The $\m$-term is given by, for example, $W\ni \kappa m_{3/2}H_uH_d$~\cite{Inoue:1991rk} which is a $Z_{4R}$ invariant since the gravitino mass $m_{3/2}$ has the $R$ charge 2. Here $\kappa$ is an $\O(1)$ constant.
The contribution to the Higgs $B$-term, $B_\mu$, is given by, e.g., $c_\m^{(2)} |F_S|^2/f_s^2$. 
The soft SUSY breaking masses, $m_{H_u}^2$ and $m_{H_d}^2,$ get contributed from $-c_{u,d}^{(2)} |F_S|^2/f_s^2.$ 
Thus, all of the soft mass parameters as well as $\m$-term in the Higgs sector can be generated. They will be treated as free parameters by considering that $\hat{c}_{u,d,\m}$ are general functions.
Although there are other contributions including the model-dependent direct couplings of the MSSM and $S$ multiplets to $|Z|^2$, we do not explicitly show them which are redundant for the soft parameters. They do not change our predictions 
as long as the couplings are $E_7$-invariant.\footnote{Our scenario is valid even if $Z$ is sequestered from all the MSSM particles and $S$. }

Note that the $\mu$-term and Higgs soft masses are of the order of $m_{3/2}$, while the pNGBs are massless at the tree-level due to the $E_7$ symmetry. Since $Z$ is charged under some symmetry, the gaugino masses are also suppressed at the tree-level. 
Therefore, radiative corrections via the $E_7$ breaking couplings, $g=\{g_1,g_2,g_3\}$ and $y=\{y_u,y_d,y_e\}$, to the SUSY particle masses are important to determine the spectra of the pNGBs and gauginos. It is well known that the anomaly mediation occurs which contributes dominantly to the masses. 
As we will show that the $S$-mediation is as important as the anomaly mediation. The contribution is determined by the $E_7$ invariance and the $S$-couplings to the Higgs multiplets.

\subsection{SUSY breaking mediation from $S$ multiplet ($S$-mediation)}

Let us first canonically normalize the kinetic term for $\phi_i$ in Eq.\,\eq{kine2} as well as the Higgs multiplets up to the $F$-term by the redefinition:
\begin{align}
\f'_i=\(1+{F_S \theta^2}/{f_s}\) \f_i, \non \\
 H_u'=\( 1+c_u F_S \theta^2/f_s\) H_u, \non \\
 H_d'=\( 1+c_d F_S \theta^2/f_s\)H_d \label{eq:frd},
\end{align}
where $c_{u}= {\partial \hat{c}_u/\partial x}$ and $c_{d}= {\partial \hat{c}_d/\partial x} $ represent the Higgs-$S$ couplings.
In the basis, the Yukawa couplings become
\beq
\eq{super}\rightarrow  \tilde{y}_u H_u' Q' u'+\tilde{y}_d H_d' Q' d'+\tilde{y}_e H_d' L' e',\eeq
where $$\tl{y}_{u}\equiv y_u(1+F_S\theta^2/f_s)^{-2}(1+c_u F_S\theta^2/f_s)^{-1},$$
$$\tl{y}_{d,e}\equiv y_{d,e}(1+F_S\theta^2/f_s)^{-2}(1+c_d F_S\theta^2/f_s)^{-1},$$ are holomorphic functions. Consequently, SUSY breaking $A$-terms are generated at the tree-level, e.g.
\beq
\laq{A}
{\cal L}\supset -(2+c_u) m_{3/2} y_u  Q' u' H'_u \equiv -A_u y_u Q' u' H'_u.
\eeq
The $A$-terms including $H_d$ also take the similar form as 
\beq
\laq{Atm}
A_{d,e}=(2+c_d) m_{3/2}.
\eeq

\paragraph{Gaugino masses}
At the loop level, the field redefinition in (\ref{eq:frd}) is anomalous and the gauge kinetic term becomes 
\beq
{\cal L}\ni \int{d^2 \theta\frac{1}{4}W_\a W^\a}\rightarrow \int{d^2 \theta\frac{1}{4} }\(1-{\frac{g^2}{4\pi^2}}\sum_{i\subset {\rm light}}{\kappa_i T_{i}} \log{\(1+ \frac{F_S \theta^2}{f_s}\)}\) W_\a W^\a \label{eq:r1},
\eeq
where the summation is taken only for light fields, we have omitted the indices of gauge (groups). 
 $\k_i=1, c_u, c_d $ for a quark/lepton, $H_u$, $H_d$ respectively, and $T_i$ is the Dynkin index for the field.
It turns out that the gaugino mass at the low energy gets contributions of 
\beq
\frac{\delta^{\rm NLS} M}{g^2} \simeq \frac{ m_{3/2}}{8\pi^2} \sum_{i\subset {\rm light}}{\k_i T_i} . \label{eq:r2}
\eeq
Note that there also exists anomaly mediation effect of $\d^{\rm AMSB} M =m_{3/2}  \beta_g/g $ with $\beta_g\equiv {d g}/{d \log{\m_{\rm RG}}}$ with $\m_{\rm RG}$ being the renormalization scale.  The form of $\delta^{\rm NLS} M$ is consistent with sigma-model anomaly mediation given in Ref.~\cite{Bagger:1999rd}. 
We again emphasize that the summations in Eqs. (\ref{eq:r1}) and (\ref{eq:r2}) only includes light pNGBs and Higgs multiplets.

Although there are various heavy modes, such as the pseudo-NG mode, $\f^a$, whose masses are much larger than $m_{3/2}$, 
these heavy modes do not affect the low energy mass spectra of the MSSM particles in our setup.
For instance, the mass term for the $\f^a, \bar{\bf 5}'$ on canonically normalized basis up to the $F$-terms is given by 
\beq
 W\ni M_{\bar{5}'} \(1+ \frac{F_S \theta^2}{f_s}\)^{-1} \(1+c_{\bar{5}'}\frac{F_S \theta^2}{f_s}\)^{-1}\f^a \bar{ 5}'_a, 
 \eeq
where $c_{\bar{5}'}$ is the coupling between $S$ and $\bar{\bf 5}'$ in the K\"{a}hler potential and $M_{\bar{5}'}$ is the SUSY mass satisfying $|M_{\bar{5}'}|\gg m_{3/2}.$ Here, we have performed field redefinitions of $\f^a$ and $\bar{5}'_a$; therefore, 
we should take into account contributions to the gaugino mass originating from the anomaly of the field redefinitions.
These contributions are given by 
\beq
\left(\frac{\d^{\rm NLS} M }{g^2}\right)_{\rm heavy}\simeq \frac{ m_{3/2}}{ 16\pi^2} (1+c_{\bar{5}'}). \label{eq:heavy_correction}
\eeq
However, the above contributions are cancelled after integrating out $\f^a$ and $\bar{5}'_a$ i.e., when the renormalization scale becomes smaller than $|M_{\bar{5}'}|$.
Threshold corrections from $\f^a$ and $\bar{ 5}'_a$ are estimated by using the formula of gauge mediation:
\beq
\left(\frac{\d^{\rm GMSB} M }{g^2}\right)_{\rm heavy}\simeq -\frac{ m_{3/2}}{ 16\pi^2} (1+c_{\bar{5}'}), \label{eq:heavy_correction2}
\eeq
which has the same form as \eqref{eq:heavy_correction}, but opposite sign.
Hence, for $\m_{\rm RG}<|M_{\bar{5}'}|$, \eqref{eq:heavy_correction} and \eqref{eq:heavy_correction2} cancel with each other: there is no contribution to the gaugino mass from $\f^a$ and $\bar{ 5}'_a$ at the low energy.
Within our setup, heavy modes generically do not contribute to the gaugino mass when the renormalization scale is lower than the mass scales of them.
This implies that, at the scale $\m_{\rm RG}\sim m_{3/2}$,
we can estimate the 
gaugino masses by taking into account the contributions only from light multiplets whose masses are less than $m_{3/2}$; therefore, the gaugino mass in (\ref{eq:r2}) is UV insensitive i.e., it is written in terms of parameters evaluated at the scale $\m_{\rm RG}\sim m_{3/2}$. Here, we have used the fact that 
$M/g^2$ is invariant under the change of the renormalization scale at the one-loop level.

\begin{figure}[t!]
\begin{center}  
\includegraphics[width=145mm]{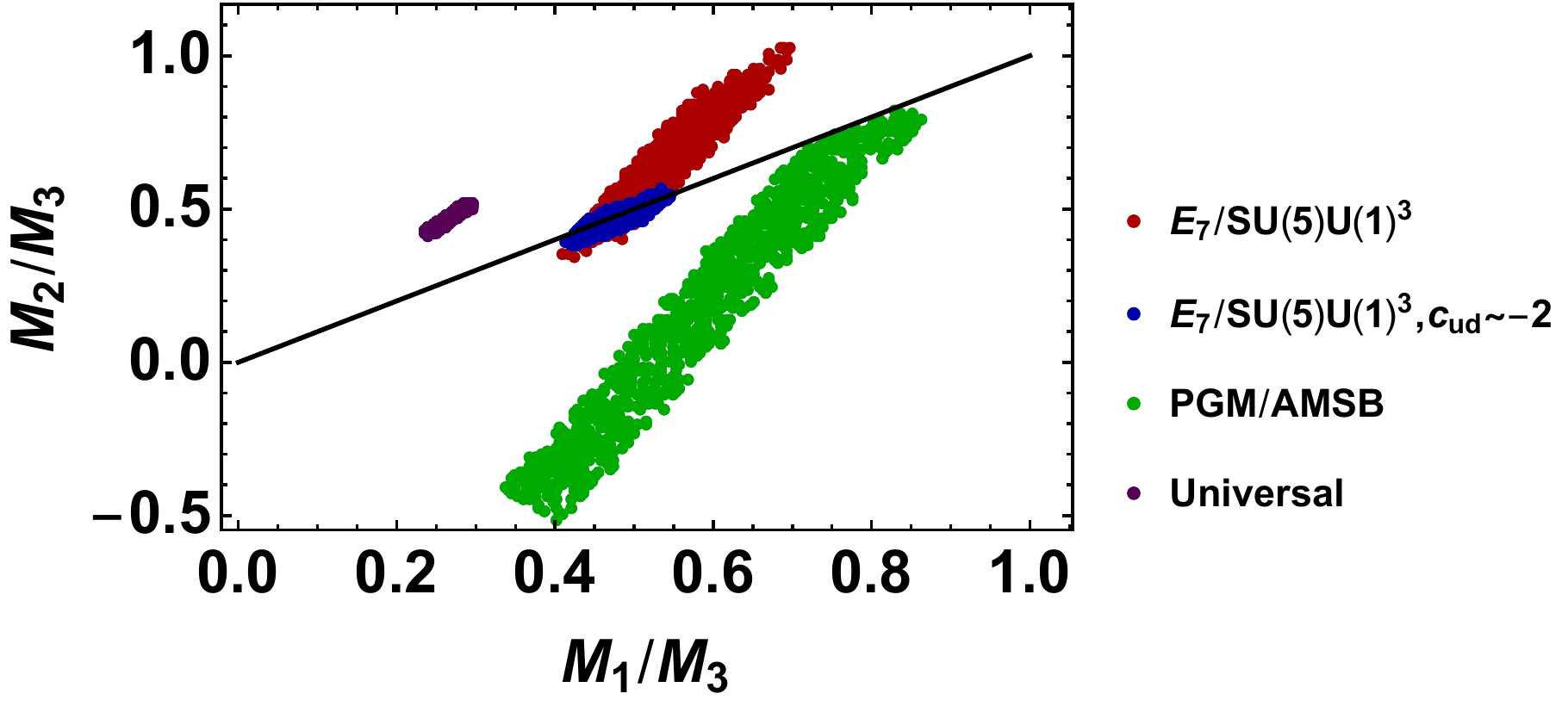}
\end{center}
\caption{The gaugino mass relation in various mediation mechanisms. The red and blue points correspond to our scenario, where we have varied $-3<c_{u, d}<3$ and $c_{u}+c_{d}=-4\pm 0.8,$ respectively. On the black solid line, $M_1=M_2.$}
\label{fig:1}
\end{figure}

Including contributions from anomaly mediation, the gaugino masses at $\m_{\rm RG}\sim m_{3/2}$ are given by\footnote{
The renormalization group running effects give only sub-leading corrections to \eqref{M1}, \eqref{M2} and \eqref{M3} because the quantity $M/g^2$ is invariant under the renormalization group evolution at the one-loop level. In numerical calculations, these corrections are included.
} 
\begin{align}
\laq{M1}
M_1&\simeq {g_1^2\over 16\pi^2}\frac{3}{5} (31+c_u+c_d )m_{3/2}, \\ \laq{M2}M_2&\simeq {g_2^2\over 16\pi^2} (13+c_u+c_d )m_{3/2},\\ \laq{M3} M_3&\simeq {g_3^2\over 16\pi^2} \, 9m_{3/2}.
\end{align}
The gaugino masses are functions of two parameters, $m_{3/2}$ and $c_u+c_d$.
In Fig.~\ref{fig:1}, we show the prediction of our scenario changing $c_u+ c_d$ from $-6$ to $6$ as red points. 
The blue points correspond to $c_u+c_d$ within $ -4 \pm 0.8 ,$ where the vacuum stability condition is satisfied (see next section).

For comparison, we also consider other typical cases: universal gaugino mass at the GUT scale leading to $M_1: M_2: M_3=g_1^2: g_2^2:  g_3^2,$ and PGM/anomaly mediation corresponding to $M_1: M_2: M_3=3/5(11+L) g_1^2: (1+L)g_2^2: -3 g_3^2$, where $L$ is $\O(1)$ parameter representing threshold corrections from Higgs-Higgsino loops.\footnote{ 
The threshold corrections also exist in our scenario, which needs a replacement of $c_u+c_d$ to $c_u+c_d+L$ in (\ref{M1}) and (\ref{M2}). However in viable regions, $L$ is suppressed as discussed later.}
The predicted values of $M_1$, $M_2$ and $M_3$ with the universal gaugino mass are shown as purple points and those in PGM/anomaly mediation are shown as green points. We vary the gauge couplings slightly taking into account the renormalization scale dependence. 
One finds that the spectra (\ref{M1}), \eqref{M2}, and \eqref{M3} are distinguishable to the other scenarios, 
that is, our scenario can be tested by measuring the gaugino masses. 

\paragraph{pNGB masses}

Now we discuss loop-induced masses for $\f_i$ by $S$-mediation.
The anomalous dimension of $\f_i$ is defined by
\beq\g_i\equiv {d \over d \log{\mu_{\rm RG}}} \log{\left[Z_i\left({|\tl{y}|^2\over \m_{\rm RG}^{2\e}}, 
\frac{\Re[\tl{g}^{-2}]^{-1}}{\m_{\rm RG}^{2\e}}
 \right)\right]}
 ,\eeq
where $Z_i$ is a wave function renormalization of $\f_i$ and we have used dimensional regularization, and $\e=2-d/2.$ Here, 
\beq
\tl{g}^{-2}\equiv g^{-2}\left[1-{\frac{g^2}{4\pi^2\m_{\rm RG}^{2\e}}}\sum_{i\subset {\rm light}}{\k_i T_{i}} \log{\(1+\frac{F_S \theta^2}{f_s}\)} \right]
\eeq  in the basis with canonically normalized gauge kinetic terms. 
The kinetic term of $\phi_i$ at the one-loop level is given by 
\beq
\laq{wfloop}
\f_i^\* e^{2g V} \f_i \(1+\frac{1}{ 2}\log{\({\m_{\rm RG}^2 \over \F_c^\*\F_c }\)} \g_i +\cdots\),
\eeq
where we have explicitly written the dependence of the compensator field~\cite{Boyda:2001nh}.
By expanding the $F$-terms, we get 
\beq
\laq{NLS}
\d^{\rm NLS}m^2_{\f_i}\simeq m_{3/2} \(\frac{\partial \g_i}{\partial \log{|{y}|^2}}  A -\frac{\partial \g_i}{\partial \log{{{g}}^2 }} \d^{\rm NLS} M \). 
\eeq
This formula can be also found from mixed modulus-anomaly mediation \cite{Choi:2004sx,Choi:2005ge,Endo:2005uy,Chowdhury:2015rja}. 
The mass squared in (\ref{NLS}) is given at $\m_{\rm RG} \sim 10^{16}\GEV$. 
Although the squark/slepton mass spectra at the scale $f_s$ may be modified by UV physics or regularization scheme (c.f. Refs.~\cite{Binetruy:2000md, Evans:2013uza}), 
we believe that the mass squares are generically loop suppressed compared with $m_{3/2}$, and the order of the masses would not be changed.
Together with the anomaly mediation, we can calculate the spectrum at the $\m_{\rm RG}\sim 10^{16}\GEV$ 
(see Appendix \ref{app2}).
For instance, the left-handed selectron mass squared is approximated as
\beq
 m^2_{\tl{e}_L}|_{\m_{\rm RG}\sim 10^{16}\GEV}\simeq -\frac{m_{3/2}^2}{(16\pi^2)^2} \(\frac{3}{2} g_2^4(13+c_u+c_d)+\frac{9}{50}g_1^4 (31+c_u+c_d)\).
\eeq
One finds that the selectron (as well as some other sfermions) is tachyonic at this scale with $c_u+c_d=\pm\O(1)$.
This is not a problem once we take into account the renormalization group (RG) running effects.\footnote{Although the tachyonic sfermions at high energy scale can introduce 
a local minimum in some direction of the scalar potential deeper than the current one, the vacuum decay rate of our universe is sufficiently suppressed after the inflation~\cite{Riotto:1995am}. 
During the inflation epoch, the sfermion fields are stabilized at the potential origin if the Higgs fields acquire large expectation values in the (approximate) flat direction $\ab{H_u}^2=\ab{H_d}^2$ due to the Hubble-induced masses. In this case the 
sfermions can have much larger masses than the Hubble parameter.   }

One of the RG effects is from the gaugino loops. The contribution is approximated as
\beq
\d^{\rm gaugino} m^{2}_{\tl{e}_L} \sim \frac{2}{16\pi^2}\(-\frac{3}{5} g_1^2 M_1^2-3g_2^2 M_2^2\) \log{\(\frac{m_{3/2}}{f_s}\)}.
\eeq
Due to large gaugino masses, \eq{M1} and \eq{M2}, and the logarithmic factor, the mass squared becomes positive at the energy scale of $m_{3/2}.$
For the first two generation squarks, the masses increase to be positive as well due to the gluino loops. 
However, the selectron and smuon masses would be below the bino or wino mass, and hence the bino or wino can not be the lightest superparticle (LSP), unless we take into account the contributions from the Higgs sector. In fact, 
the slepton LSP conflicts with the standard cosmology.\footnote{One may avoid this by taking the higgsino mass $|\m|$ much smaller than $m_{3/2}$, so that the higgsino is the LSP. Also, one may assume an $R$-parity violation in which case the dark matter may be $\Im[S]$. In these cases, the gaugino mass pattern may be tested 
from the decays of the gauginos to the LSP at  the collider experiments. In this paper, however, we do not take these further assumptions.} 

\vspace{12pt}
In the Higgs sector, the soft breaking masses exist at the tree-level, which can induce sizable effects on the squark and slepton masses through the RG running. 
The Higgs soft masses are given by 
\beq
\laq{Higgsmass}
m_{ H_u}^2\simeq m_{ H_d}^2 \simeq -c_h m_{3/2}^2,
\eeq
where we have taken $m_{ H_u}^2\simeq m_{ H_d}^2$ so that the $s$-term, $s= \sum_{i\subset {\rm light}}{ Y_i m^2_i}$ with $Y_i$ being the hypercharge, is vanishing.
In this case,  
\beq c_u = c_d \laq{cucd}\eeq
should also hold.\footnote{ On the other hand, $c_u \sim c_d \sim -2$ will be forced by the vacuum stability.}
 This is because the canonical normalization of $H_{u,d}$ induces the contribution to $s$ proportional to $c_u^2-c_d^2.$
We emphasize that $c_h$ can be positive, which leads to the RG effects via the Higgs loops, called Higgs mediation~\cite{Yamaguchi:2016oqz, Yin:2016shg,
Yanagida:2016kag,
Yanagida:2018eho}. The negative and large $m_{H_u}^2 +A_t^2,m_{H_d}^2 +A_d^2$ dominantly contribute to the squark and slepton masses of the third generation through the one-loop RG running~\cite{Yamaguchi:2016oqz}. For instance, 
the left-handed stop mass squared gets
\beq
\laq{HM}
\d^{\rm HM} { m}_{{{\tl{t}_L}}}^2 \sim {2 \o 16 \pi^2} m_{3/2}^2  \(y_{t}^2\(-c_h +(2+c_u)^2\)  +y_{b}^2 \(-c_h +(2+c_d)^2\)\)\log{\({m_{3/2}\o f_s}\)} ,\\
\eeq
where $y_{t}\AND y_{b}$ are the Yukawa couplings of top and bottom quarks, respectively. 
The contribution from Higgs mediation to the third generation squarks and sleptons can be as large as $\O(10)\% m^2_{3/2}$ for $c_h-(2+c_u)^2=\O(1).$ The one-loop contributions to the first two generations are highly suppressed by the small Yukawa couplings. Their masses are dominantly generated at the two-loop level~\cite{Yin:2016shg}. For instance, the selectron mass obtains a two-loop contribution of 
\beq
\laq{HM2}
\d^{\rm HM} { m}_{{{\tl{e}_L}}}^2 \sim -\frac{6g_2^4}{( 16 \pi^2)^2} m_{3/2}^2 c_h\log{\({m_{3/2}\o f_s}\)} ,\\
\eeq
which makes the selectron heavier than the bino or wino. Then, the neutralino becomes the LSP. 

Notice that the negative and large $m_{H_u}^2$ and $m_{H_d}^2$ do not mean that the quadratic terms in the Higgs potential are negative. 
This is because they appear as the combinations of $m_{H_u}^2+\mu^2$ and $m_{H_d}^2+\mu^2$ in the potential with large $\ab{\mu}$. In this case, $\tan\b\gtrsim \O(10)$ is required for a successful EW symmetry breaking. 
This can be found from the non-tachyonic condition of the MSSM Higgs boson at the $\mu_{\rm RG}\sim m_{3/2},$
\beq
\laq{Acon}
m_A^2\simeq m_{H_d}^2-m_{H_u}^2>0,
\eeq
which differs from the tree-level condition \eq{Higgsmass}. Thus, 
the radiative corrections to the mass differences are needed. 
 The radiative corrections are estimated by the RG equations:
$
\frac{d}{d\log\mu}m_{H_u}^2\simeq \frac{1}{16\pi^2} 6y_t^2 m_{H_u}^2,~~ \frac{d}{d\log\mu}m_{H_d}^2\simeq \frac{1}{16\pi^2} (6y_b^2+y_\tau^2) m_{H_d}^2.
$
This implies the inequality \eq{Acon} can be satisfied only when $y_\t,y_b\gtrsim y_t$, i.e.  $\tan\b\gtrsim \O(10)$. (Notice $m_{H_u}^2,m_{H_d}^2<0$.) The RG running effects will be taken into account by solving two-loop RG equations in the next section.

\label{sec:m2}

\section{Phenomenological consequences and miracle}
Now let us perform numerical calculations for SUSY mass spectra using {\tt SuSpect\,2.4.3}~\cite{Djouadi:2002ze} with modifications. The conditions, \eq{A}, \eq{Atm}, (\ref{M1})-(\ref{M3}), and \eq{cucd},
 together with the scalar masses in Appendix \ref{app2} are set at the input scale $M_{\rm inp}=10^{16}\GEV.$
\subsection{Constraints and parameter region}

We notice again that the squarks and sleptons are massless at the tree-level, which implies that we may have too large trilinear-terms, e.g. \eq{A}.
Although the squarks and sleptons can get masses at the loop-levels, this may induce a true 
minimum with color/charge breaking  and cause rapid decay of the electroweak vacuum. 
In particular, a larger trilinear Higgs-stop coupling would result in a deeper true vacuum and a smaller potential barrier. 
Here, we show that a consistent EW vacuum requires $c_u=c_d\sim -2$ so that the size of the $A$-terms are small enough according to \Eqs{A} and \eq{Atm}. This leads to a surprising prediction: quasi-degenerate masses of the bino and wino.

\begin{figure}[t!]
\begin{center}  
\includegraphics[width=125mm]{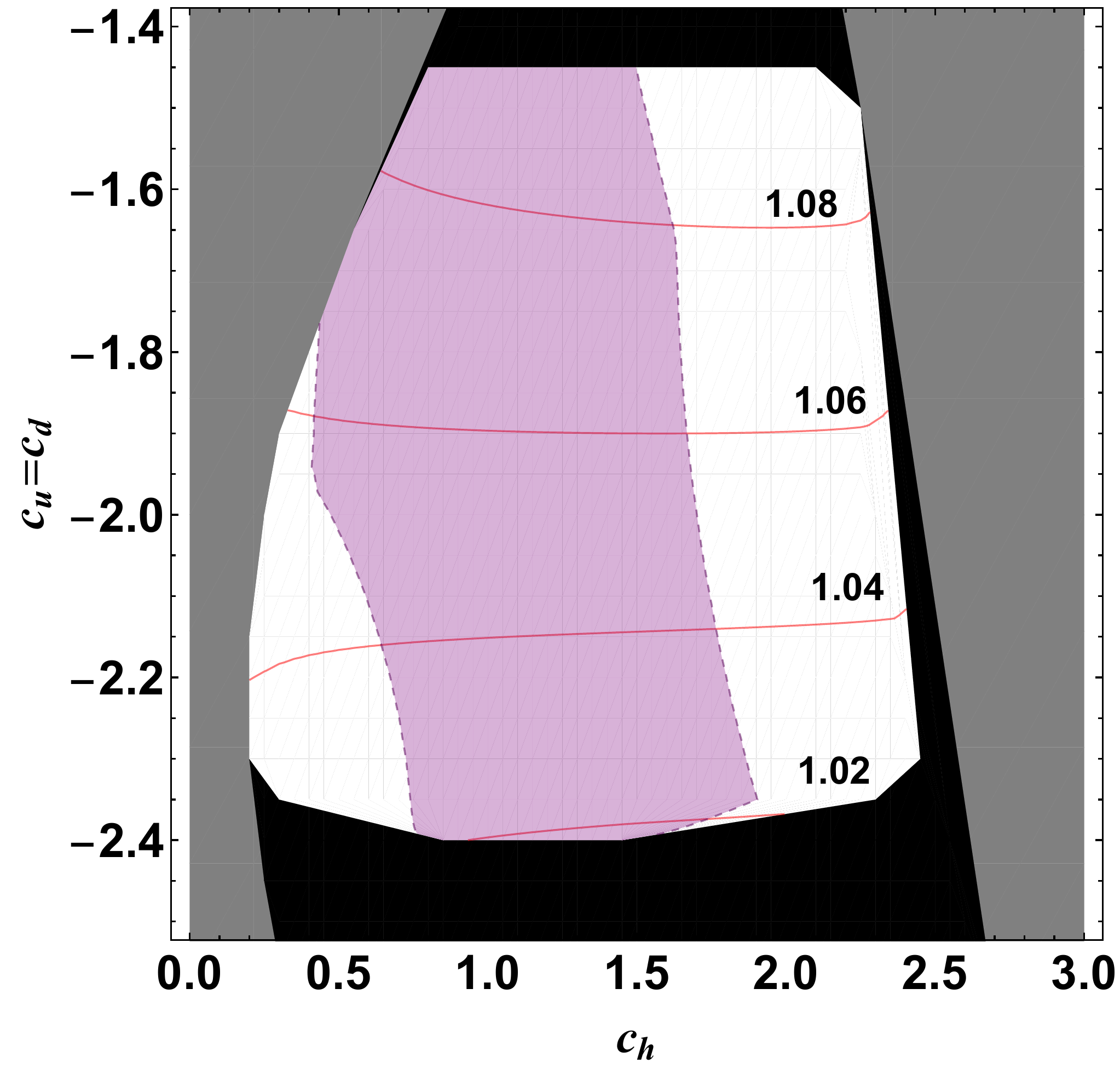}
\end{center}
\caption{The viable parameter region for $m_{3/2}=60\TEV$ and $\tan\b=50$ on $c_h$-$c_u(=c_d)$ plane. We fix ${\rm sign}{[\m]}=-1.$ The contours denote the mass ratio of the lightest chargino to the lightest neutralino. The black region contradicts with \eq{VC}. The gray region is excluded due to a tachyonic/too-light scalar particle.   On the purple range the lightest neutralino is the LSP, which is bino-like.  Left (right) to the purple range, the LSP is the slepton (squark). 
Throughout the paper, we take the top mass as $M_t=173.21\GEV$ and QCD coupling constant as $\alpha_s(m_Z)=0.1181.$ }
\label{fig:vs}
\end{figure}

\begin{figure}[t!]
\begin{center}  
\includegraphics[width=125mm]{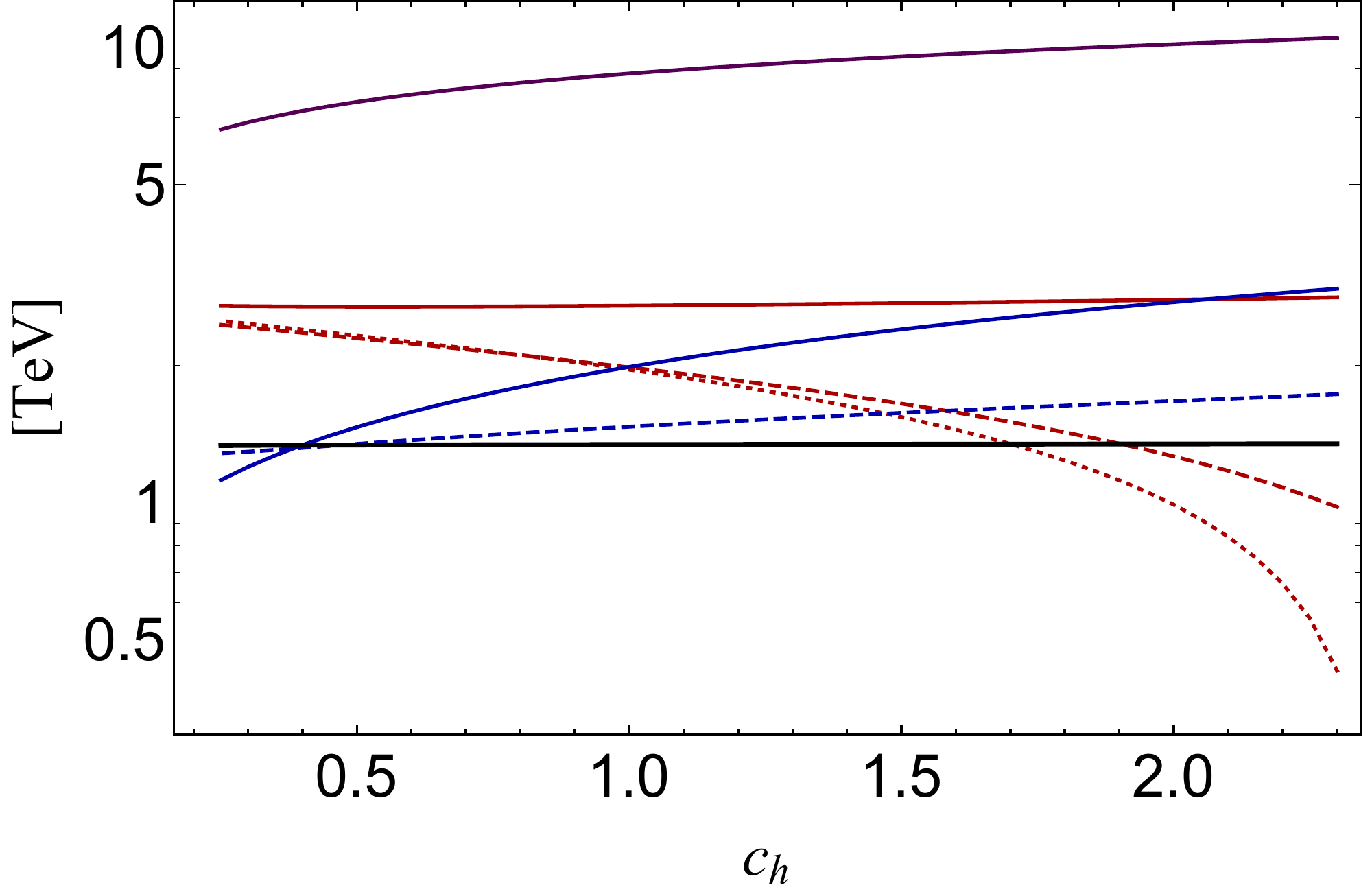}
\end{center}
\caption{ Sfermion mass spectra at $c_u=c_d=-2$ by varying $c_h.$ Other parameter sets are same as in Fig.\,\ref{fig:vs}.  The purple line is the SUSY scale $\sqrt{m_{\tilde{t}_R}m_{\tilde{t}_L}}.$ The red solid (dashed, dotted) line represents $m_{\tl{u}_L}(m_{\tl{u}_R},m_{\tl{d}_R})$. 
The blue solid (dashed) line shows $m_{\tl{e}_L}(m_{\tl{e}_R})$. 
The horizontal black solid line shows the bino-like neutralino mass.   }
\label{fig:sfe}
\end{figure}

To discuss the vacuum (meta)stability, we adopt the empirical constraint for the squarks-Higgs system from Ref.~\cite{Kusenko:1996jn}
\beq
\laq{VC}
7.5(m_{\tl{Q}_3}^2+m_{\tl{u}_3}^2)>3\mu^2 +A_t^2.
\eeq
In Fig.~\ref{fig:vs}, we show the parameter region satisfying this bound on $c_h$-$c_u (=c_d)$ plane for $m_{3/2}=60\TEV, \tan\b=50$ and $\m<0$. 
The black (gray) region is excluded by the vacuum stability constraint (tachyonic/too light scalars).  
On the purple shaded region the neutralino is the LSP. 
The contour represents the ratio of the wino-like chargino to the bino-like neutralino. 
Firstly, one finds that the vacuum stability requires that  \beq c_u= c_d =-2 \pm \O(0.1).\laq{viab}\eeq
Secondly, the region with a neutralino LSP, which is bino-like, exists due to the RG running effects from the Higgs soft masses. The numerical result of the sfermion mass dependence on $c_h$ is shown in Fig.~\ref{fig:sfe} by fixing $c_u=c_d=-2.$ 
One finds that the selectrons (blue lines) become heavier than the bino-like neutralino LSP (black line) with $c_h\gtrsim 0.4.$ This is consistent with (\ref{HM2}).
Lastly, a surprising fact is that the masses of the bino-like neutralino and the wino-like chargino are almost the same when \eq{viab} is satisfied.
The masses at most differ by $\O(1)\%$. 
The physics of this coincidence will be discussed soon. 

Let us comment on possible corrections to the mass difference of the bino and wino, and 
show that the degeneracy should not be removed by higher order corrections. As we noted, the formula of \eq{M1}-\eq{M3} is insensitive to the UV physics up to one-loop level but there are model-dependent UV corrections and RG running effects at the two-loop level or higher. They contribute to the gaugino masses and thus to the bino-wino mass difference by $\O(1)\%.$ 
Other contributions are threshold corrections from MSSM particles, which we have included in the numerical calculations (see Ref.~\cite{Pierce:1996zz} for the dominant threshold corrections). 
These contributions are also at most $\O(1)\%$ level.  
In particular, contributions from Higgs-Higgsino loops are suppressed because $\tan\b\gtrsim \O(10)$~\cite{Yanagida:2016kag}.\footnote{A possibility we did not consider here is that the multiplets $\f_a, \bar{\bf 5}'$ are lighter than $m_{3/2}.$ 
In this case, there should be also corrections to the mass difference up to around $1$-$10\%$, depending on $c_{\bar{5}'}$. 
}

\begin{figure}[t!]
\begin{center}  
\includegraphics[width=125mm]{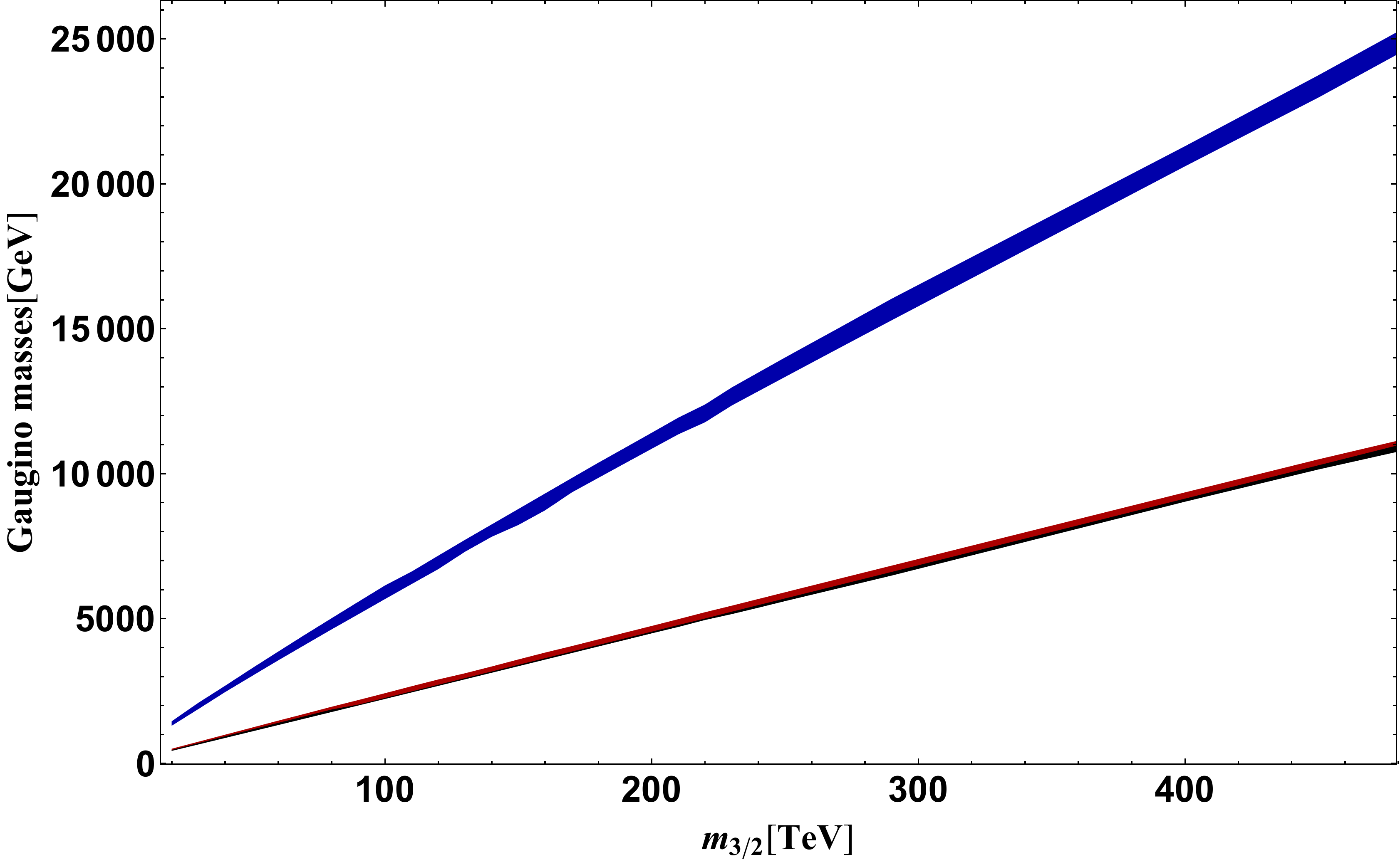}
\end{center}
\caption{The masses of gluino (blue band), lightest chargino (red band) and lightest neutralino (black band) from top to bottom. 
We vary $40 \leq \tan\b \leq 80, 0.01\leq c_h\leq3$ and  $-2.7\leq c_u=c_d\leq -1.3$ with ${\rm sign}{[\m]}=-1$ fixed. Only data points with the (meta)stable electroweak vacuum and neutralino LSP are shown. }
\label{fig:DM}
\end{figure}

The gaugino masses, \eq{M1}-\eq{M3}, are almost fixed by one free parameter $m_{3/2}$ due to \eq{viab}, which leads to a unique gaugino mass spectrum. 
The mass dependences of gluino (blue band), the lightest chargino (red band), and the lightest neutralino (black band at the lower boundary of the red band) on $m_{3/2}$ are shown in Fig.\,\ref{fig:DM}. We vary $\tan\b=40$\,-\,$80$, $c_h=0.01$\,-\,$3$, $-2.7<c_u=c_d<-1.3$ with ${\rm sign}{[\m]}=-1$. These ranges of the parameters will be used in the following numerical calculations. 
This mass relation of gauginos is one of the robust predictions in our scenario.

 \begin{figure}[t!]
\begin{center}  
\includegraphics[width=125mm]{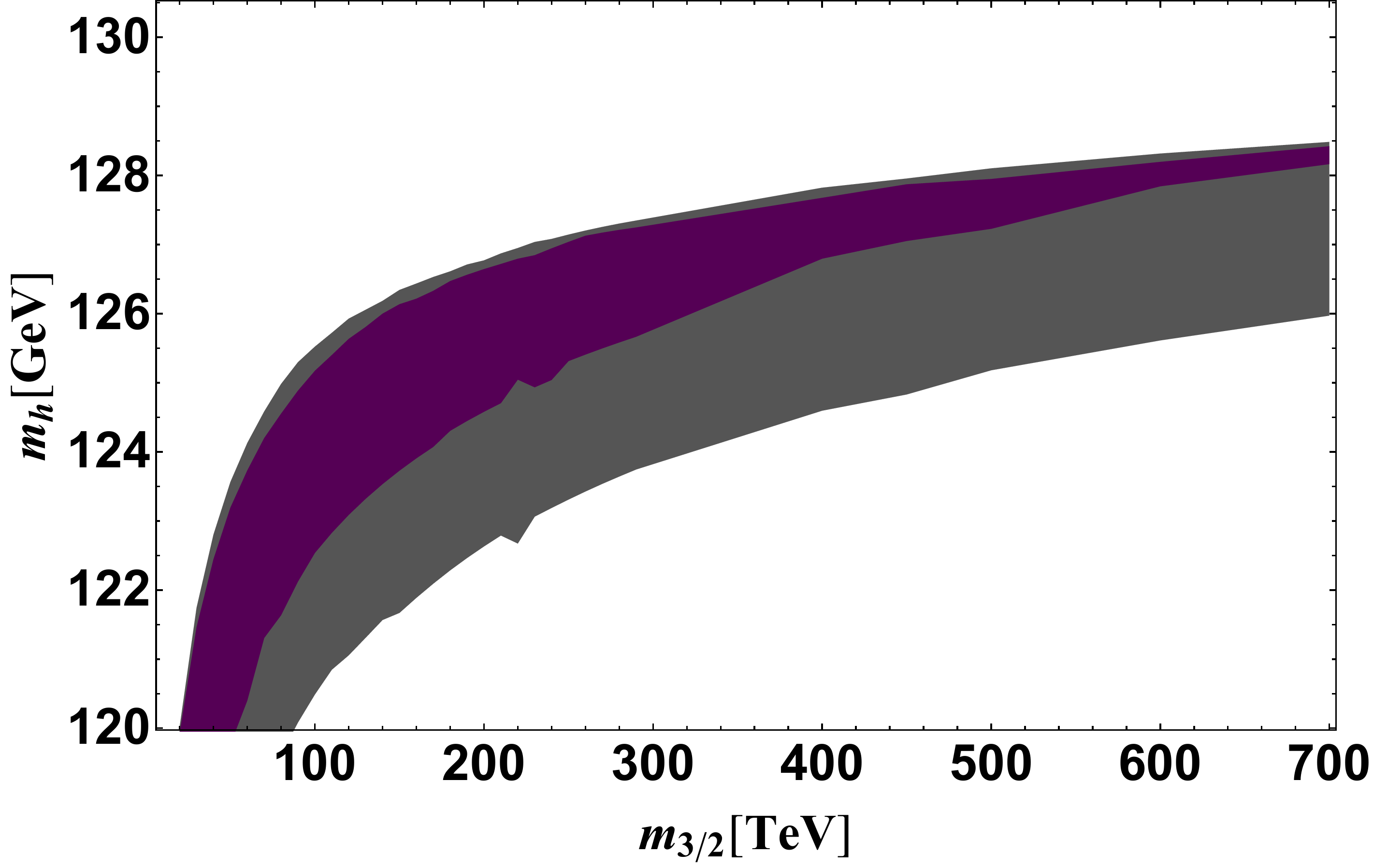}
\end{center}
\caption{The Higgs boson mass for the neutralino LSP region (purple band) and the region with vacuum stability (gray band). The ranges of the parameters are same as in Fig.\,\ref{fig:DM}.   }
\label{fig:2}
\end{figure}

Another constraint is the Higgs boson mass $\simeq 125\GEV \pm 3\GEV$. To estimate the Higgs boson mass we use {\tt SUSYHD\,1.0.2}~\cite{Vega:2015fna}. The plot is given in Fig.~\ref{fig:2}.
The region with the neutralino LSP corresponds to the purple band. 
The gray band represents the whole region consistent with \eq{VC}. 
It turns out that the Higgs boson mass predicts
\beq
\laq{Higgs}
m_{3/2}\simeq 40\TEV-600\,{\rm TeV}.
\eeq 
We notice that the prediction on the Higgs boson mass is slightly smaller than the case with smaller $|\mu|$-term. (See also Tab.\,\ref{tab:2} and Ref.\,\cite{Draper:2013oza})This is due to the negative sbottom-loop contribution on the quartic coupling of the 
Higgs boson~\cite{Vega:2015fna}. 

\subsection{Bino-wino coannihilation} 

The scenario predicts that the bino and wino have almost degenerate masses, and the bino-like neutralino is the LSP in most cases.
When the bino-like neutralino mass is smaller than $\sim 3\TEV$ and the mass difference between the wino and bino is around or below $\O(10)\GEV,$ the dark matter abundance can be explained through the bino-wino coannihilation, taking into account the Sommerfeld effect for the wino~\cite{Baer:2005jq,ArkaniHamed:2006mb,Ibe:2013pua,Harigaya:2014dwa, Duan:2018rls}.\footnote{When $c_h$ is parametrically large, the squark mass can be close to the LSP mass due to the two-loop RG effects.  In this case, the squark-wino-bino coannihilation takes place and the mass of the LSP larger than $\sim 3\TEV$ may still be allowed.} 
The required mass difference for the correct relic abundance can be explained for $c_u=c_d \lesssim -2$.

The range of the gravitino mass consistent with the thermal dark matter~\cite{Harigaya:2014dwa}, $M_1\lesssim 3\TEV$, and the Higgs boson mass \eq{Higgs} is
\beq
40\TEV \lesssim  m_{3/2}\lesssim 150\TEV,
\eeq
which corresponds to our viable region. The mass difference between bino and wino can be small enough to have correct abundance with small enough $c_{u,d}$. 
Although the bino dark matter gets a correct abundance via the coannihilation with wino, the dark matter today has a suppressed interaction rate with the SM particles. Therefore, the direct and indirect detections of dark matter are  difficult, i.e. our scenario is almost free from the constraints.

\subsection{Yukawa and gauge coupling unification}

\begin{figure}[t!]
\begin{center}  
\includegraphics[width=75mm]{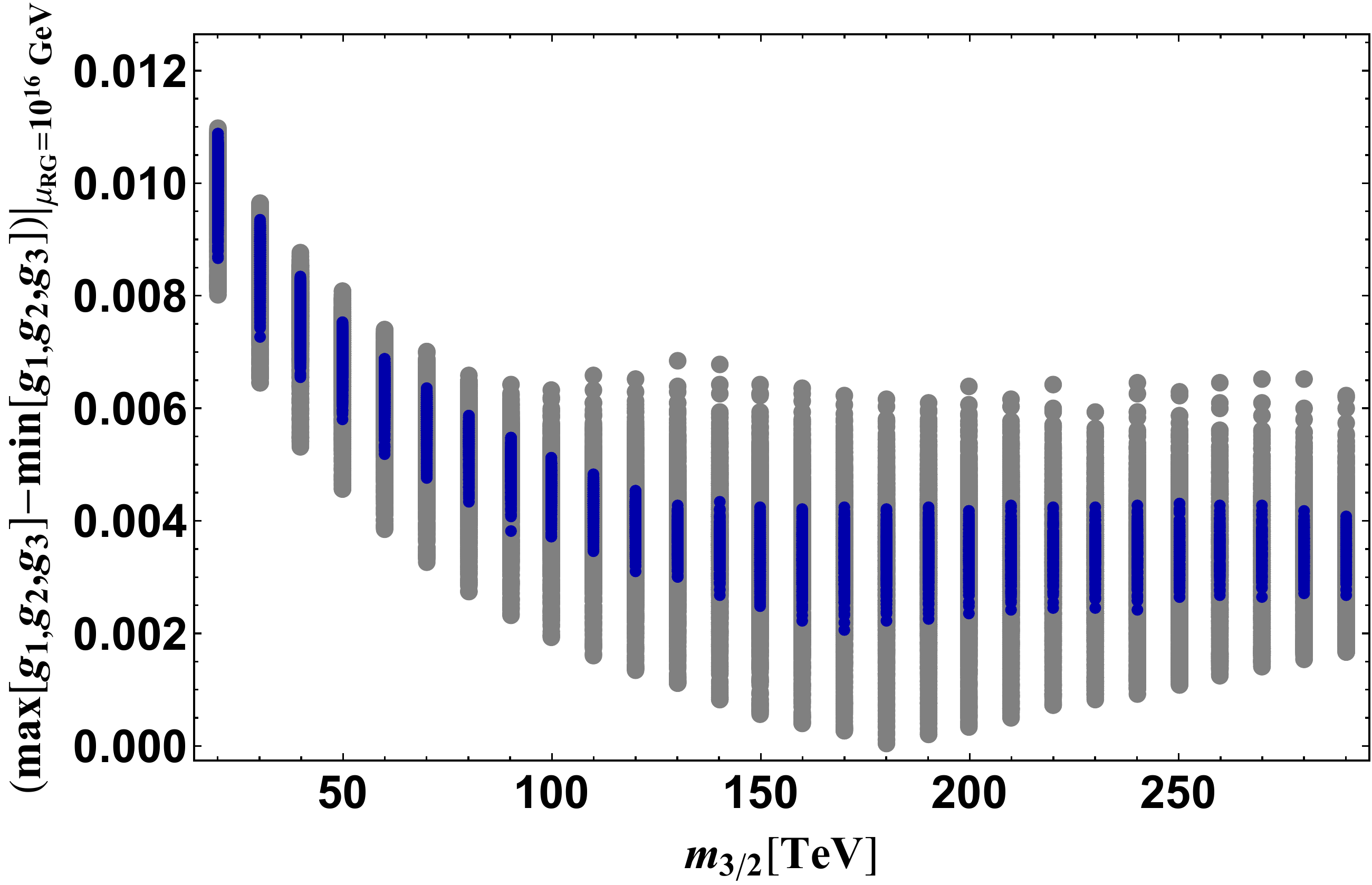}
\includegraphics[width=72mm]{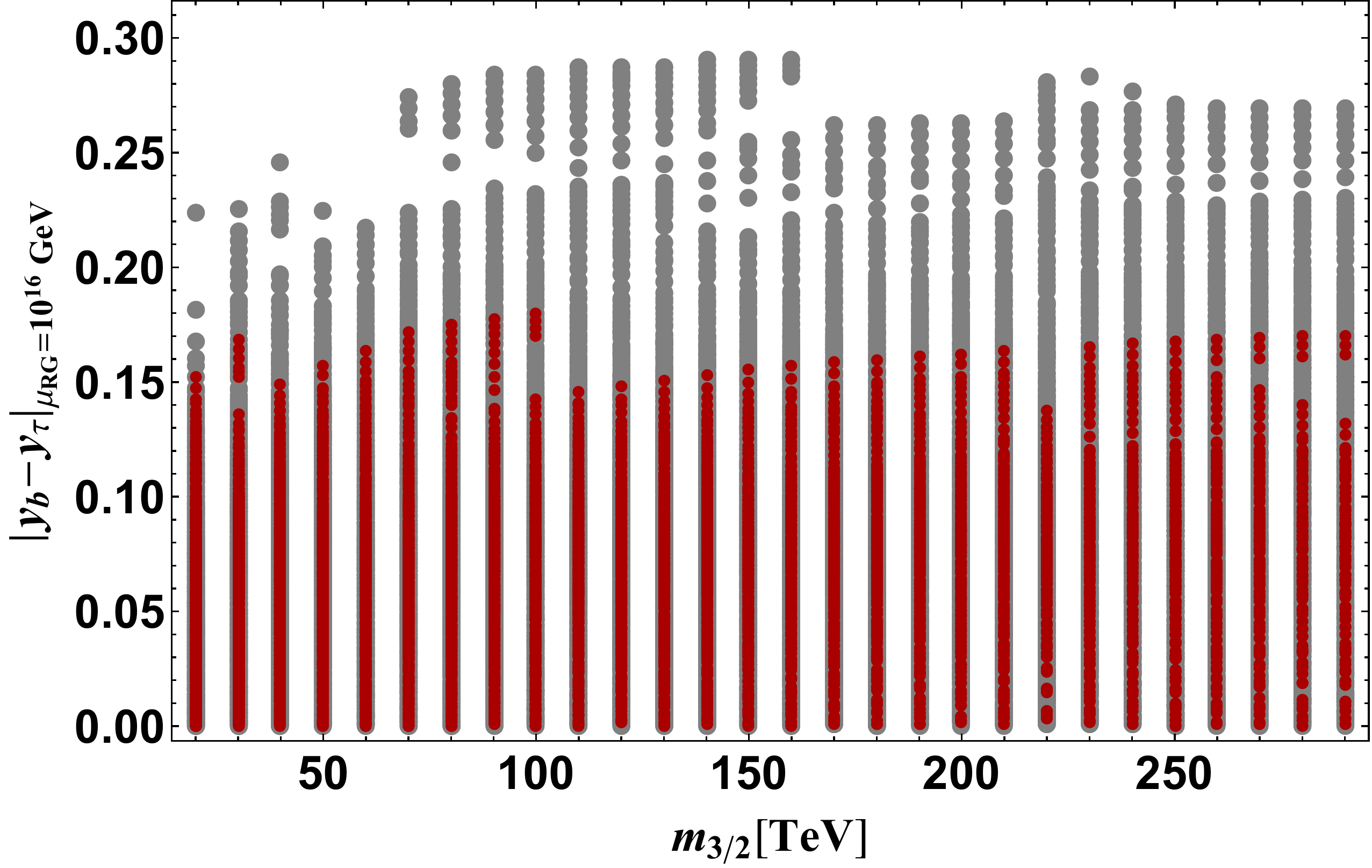}
\end{center}
\caption{Precisions of gauge and Yukawa coupling unifications in the left and right panel, respectively. 
The gray points satisfy \,\eq{VC}. On blue/red points, the neutralino is LSP. The ranges of the parameters are same as in Fig.\,\ref{fig:DM}.   }
\label{fig:yukawa}
\end{figure}

Since the down-type quarks and left-handed leptons in the MSSM form complete $SU(5)$ multiplets $\bar{\bf 5}$,
the Yukawa couplings, $y_{d}$ and $y_{e}$, are expected to be unified at the GUT scale. As we have discussed, to have neutralino LSP, the Higgs mass squares should be negatively large, and thus $\tan\b\gtrsim \O(10)$ is required from \eq{Acon}. This is a good condition for the $b-\t$ Yukawa coupling unification. 
The difference between $y_b$ and $y_{\tau}$, $|y_b-y_\tau|$ at the renormalization scale $\m_{\rm 
RG}=10^{16}\GEV$, is plotted in Fig.\,\ref{fig:yukawa} (right panel) with respect to the $m_{3/2}$. 
We also show the precision of the gauge coupling unification, $\max{(g_1, g_2, g_3)}-\min{(g_1, g_2, g_3)}$ (left panel).  
On the gray data points, the vacuum stability constraint is satisfied. On the blue and red points, the LSP is the lightest neutralino. 
We find that, when the neutralino is the LSP, 
the gauge and Yukawa coupling unifications can occur at $\O(1)\%$ precision level. (The analysis of the coupling unification is based on Ref.~\cite{Yanagida:2018eho}.)
The $\O(0.01)$ difference may come from the threshold corrections from GUT theory which may be the order $\O(\vev{\Phi_{GB}}/M_P) \OR \O(1/16\pi^2)$. Here $\vev{\F_{GB}}\sim 10^{16}\GEV$ is the VEV of a GUT breaking field. 
We do not care about the unifications of the Yukawa couplings for the first two generations, because the threshold corrections may be dominant. 

Notice that the $f_s$ can not be much smaller than $10^{16}\GEV$ for the perturbative gauge coupling unification.  This is because otherwise higher dimensional terms, e.g. the last term of \eq{kine2}, become strongly-coupled below the GUT scale.

\subsection{Miracle and collider signatures}
\begin{table*}[!t]
\caption{\small Sample data points. The model parameters except for $\tan\b$ are set at $M_{\rm inp}=10^{16}\GEV.$ All the points give the correct dark matter abundance $\Omega_{\chi_1} h^2\sim 0.1$ according to Ref.\,\cite{Harigaya:2014dwa}.}
\label{tab:2}
\begin{center}
\begin{tabular}{|c||c|c|c|}
\hline
 Parameters & {\bf I} & {\bf II}  & {\bf III}   \\
\hline
$m_{3/2} $[TeV]& 60 &  70 &  $130$ \\
${\tan\b}$ & 53 & 49  &  50\\
$c_h$ & 0.84 & 1.7& 0.6  \\
$c_{u}=c_d$ & -2.4 & -2.3& -1.8  \\
$\delta^{\rm axion} M_{1,2,3}/m_{3/2}$ & 0 & 0& -0.003  \\
\hline 
 spectrum & [TeV] & [TeV]  & [TeV]  \\ 
\hline
$\chi_{1}$   & 1.33 & 1.56& 2.74\\
$\chi_{2}$    & 1.36 & 1.59&  2.74\\
$\tl{g}$  & 3.81 & 4.38 & 6.64\\
\hline
$\chi_{3,4}$  & 48 &79  &87\\
$\tl{u},\tl{c},\tl{d},\tl{s}$ & 2.2-2.8&1.7-3.3 &3.7-4.9\\
$\tl{t},\tl{b}$ & 21-22&32-36  &40-42\\
$\tl{\m},\tl{e}$ & 1.5-2.3 & 1.7-3.6&2.8-4.3\\
$\tl{\tau}_{1,2}$ & 15,22 &23,33&27,38\\
\hline
$h_{\rm SM}$ [GeV] &122& 124 & $124$ \\
\hline
couplings at $\mu_{\rm RG}=$& $10^{16}\GEV$& $10^{16}\GEV$  &$10^{16}\GEV$ \\
\hline
$y_b$  &$0.58$& $0.42$ & $0.58$ \\
$y_\tau$  &$0.65$& $0.51$ & $0.60$ \\
$g_1,g_2,g_3$  &$\simeq 0.69$& $\simeq 0.69$ & $\simeq 0.69$ \\
\hline
\end{tabular}
\end{center}
\end{table*}

Let us summarize what we have shown. 
We consider the $E_7/\SU(5)\times \U(1)^3$ NLS model coupled to supergravity with the dynamical scale, $10^{16}\,{\rm GeV} \lesssim f_s\ll M_P.$
To break SUSY, we have introduced the SUSY breaking field charged under some symmetry, 
avoiding the Polonyi problem.\footnote{This assumption was also important for the prediction of the quasi-degenerate bino and wino since otherwise $Z W_\a W^\a$ is allowed. The term significantly changes the gaugino masses at the tree-level.}
To make the model realistic, $E_7$ is explicitly broken by the gauge interactions, and the Yukawa interactions among the Higgs (non-NG) multiplets and NG multiplets. Then, massless gauginos, sleptons and squarks at the tree-level but non-vanishing Higgs soft mass parameters are predicted.

At the loop level, 
$S$-mediation as well as anomaly mediation gives non-vanishing gaugino masses.
This is because the $S$ must have couplings to the NG multiplets as well as the compensator field, which are determined by the $E_7$ symmetry. 
The contributions from $S$-mediation and those from anomaly mediation are almost UV-insensitive, 
and thus the gaugino mass relations, \eqref{M1}-\eqref{M3}, are justified at $\m_{\rm RG}\sim m_{3/2}.$
By searching for a consistent electroweak vacuum, as any scenario does, 
we found that the mass difference of the bino and wino is within $\O(1\%)$. 
Note that the mass degeneracy is essentially determined by the observables at the low energy: the gauge coupling constants and the structure of the EW vacuum.  
Therefore, our prediction is highly non-trivial and robust. 
As a natural consequence, we can have a successful thermal bino dark matter with bino-wino coannihilation. 
From conditions for the dark matter abundance and the Higgs boson mass we get the gravitino mass of 
\beq
\laq{m32}
40\TEV\lesssim m_{3/2}\lesssim 150\TEV. 
\eeq
Moreover, we found that the Yukawa coupling unification and gauge coupling unification occur at the level of $\O(1\%)$ in most of the viable region.   
The gravitino heavier than $\mathcal{O}(10)$\,TeV decays well before the epoch of the Big Bang nucleosynthesis (BBN); therefore, the gravitino problem is alleviated. The sample data points in the miraculous region are given in Table.\,1 (see the next section for $\delta^{\rm axion} M_{1,2,3}/m_{3/2}$). All the points have the consistent dark matter abundance from the bino-wino coannihilation.\footnote{We have neglected the coannihilation effect between the bino and the slepton/squark.}

Now let us discuss collider signatures. The lower bound of \eq{m32} corresponds to the LSP mass around $1\TEV$. 
This is consistent with null results of LHC SUSY searches (up to now) 
because the LSP is quite heavy: even if SUSY particles are produced, their decays to the LSP are difficult to be identified.
In the future, the scenario can be tested by collider searches for the gauginos, especially the 
gluino. 
The existence of the gluino can be checked at the LHC up to $\sim 3\TEV$.  The gluino can be searched for up to $\sim 6\TEV$ and $\sim 12\TEV$ at future $33\TEV$ and $100\TEV$ collider experiments, respectively~\cite{Cohen:2013xda}. In particular, the $100\TEV$ colliders can cover the whole viable region.\footnote{On the other hand, it was shown that the wino up to mass $2.3\TEV$ may also be tested by precise studies on pair productions of charged leptons or of a charged lepton plus a neutrino at future $100\TEV$ collider experiments~\cite{Abe:2019egv}. }
Here, we have assumed that the squarks and sleptons are heavy enough for a conservative purpose.\footnote{The model-dependent UV corrections may increase the squark and slepton masses.
However, these corrections do not change the predictions of the bino-wino coannihilation and the unification of the bottom and tau Yukawa couplings
significantly. The soft parameters relevant to the predictions are either in the gaugino sector or the Higgs sector. The former parameters were shown to be irrelevant to the UV corrections, and the latter parameters are at tree-level whose contributions via the one-loop RGE  should dominate over the UV radiative corrections.  } 
The light squarks and sleptons, which appear in our numerical estimations, make our scenario easier to be tested due to the enhanced production cross section of the SUSY particles. In fact, the first two generation squarks 
are lighter than the gluino in the viable region. 
In that case, $m_{3/2}\lesssim 80\TEV  \,(180\TEV,  420 \TEV)$ may be tested at the LHC ($33\TEV$, $100\TEV$ colliders)~\cite{Cohen:2013xda}. 
This gives an optimistic view to the collider searches: the LHC can test a large fraction of the viable region and a 33$\,$TeV collider may cover the whole region.

Before ending this section, let us comment on three things. 
One is the existence of the region consistent with the unification of top-bottom-tau Yukawa couplings. 
The other is the region explaining the muon $g-2$ anomaly. 
Although we have focused on the region with ${\rm sign}[\m]<0$, where SUSY contributions to the muon $g-2$ are negative, 
the muon $g-2$ anomaly may be explained for ${\rm sign}[\m]>0$ with a small enough $m_{3/2}.$ 
The shortcoming, in this case, is that the bottom-tau Yukawa coupling unification is difficult to be explained.   
The last thing is the little hierarchy between the stop mass scale and the weak scale.  In order to make the slepton heavier than the neutralino, the Higgs loop effect is needed. This effect also increases stop masses (see \Eqs{HM} and \eq{HM2}) so that the Higgs boson mass can be 
explained. However, this implies that the stop mass is much larger than the weak scale.

\section{$\Im[S]$ as the QCD axion}
\label{sec4}
In our scenario, 
there is another massless particle, the scalar of $\Im[S]$, due to the shift symmetry \eq{1}.
An interesting possibility is that $\Im[S]$ is a QCD axion. 
In this case, $S$ couples to the field strength superfields of $\SU(5)$ as
\beq
\laq{ac}
{\cal L}\supset \int{d^2\theta (c_s g_5^2) \frac{S}{ M_*} W_\a W^\a } + h.c., 
\eeq
where $c_s$ is a real constant, $g_5$ is gauge coupling of $SU(5)$, $M_*$ is a cut-off scale, and $W_{\alpha}$ is the field strength superfield. Here, we take the canonically normalized kinetic term for the gauge fields.
Notice that (\ref{ac}) breaks the shift symmetry (\ref{1}) explicitly.
Since the $\SU(3)_c$ coupling becomes strong at around the QCD scale, the $\Im[S]$ acquires a non-flat potential and eliminates the strong CP phase at the vacuum. 
The decay constant of the QCD axion is given by
\beq
f_a= \frac{1}{32\pi^2}\frac{M_*}{ |c_s|}.
\eeq
Due to the $F$-term \eq{F}, the gaugino masses get additional contributions of 
\beq \delta^{\rm axion} M_i \simeq -  \frac{g_i^2}{16\pi^2} \frac{f_s}{f_a} {\rm sign}{[c_s]} m_{3/2}. \eeq
For $(f_s/f_a) \lesssim \mathcal{O}(1)$,  
the above contribution to the mass difference between bino and wino is at most $\O(1)\%$, and does not spoil the successful explanation of the relic dark matter abundance.
There is also constraint on the decay constant of the QCD axion from the black hole super-radiance, $f_a\lesssim 10^{17}\GEV$~\cite{Arvanitaki:2014wva,
Cardoso:2018tly,
Stott:2018opm}. These facts imply $M_*\sim 10^{18}\mathchar`-10^{19}$\,GeV, i.e. $M_* \sim M_P$ for $c_s=\mathcal{O}(1)$. (Remember that $f_s \gtrsim 10^{16}$\,{\rm GeV}. )

The axion abundance is overproduced if the initial amplitude of the axion coherent oscillation is set to be $\O(f_a)$. This overproduction problem could be absent by taking the initial amplitude smaller than $O(10^{-3}) f_a$.\footnote{This may be due to an anthropic selection. On the other hand, a small amplitude can be obtained if the Hubble parameter during (eternal-)inflation is smaller than the QCD scale~\cite{Graham:2018jyp, Guth:2018hsa, Ho:2019ayl}. Low-scale inflation can be possible with successful reheating if the inflaton is an axion-like particle~\cite{Daido:2017wwb, Daido:2017tbr, Takahashi:2019qmh}.} 
The axion may be tested by the further spin measurements of the black holes.

The axion can compose a dominant fraction of dark matter, when the neutralino LSP abundance is not enough.
The axion dark matter can be tested by the ABRACADABRA experiment~\cite{Ouellet:2019tlz} or a dark matter radio wave experiment~\cite{Silva-Feaver:2016qhh}. (See also Ref.~\cite{Irastorza:2018dyq} for review and Ref.~\cite{Salemi:2019xgl}  for the latest result of the ABRACADABRA  experiment.)

Lastly, we mention the moduli problem caused by the scalar component of $\Re[S].$ The $\Re[S]$ during the inflation may not be at the potential minimum. After the inflation it starts to oscillate around the potential minimum and the energy density of the coherent oscillation easily dominates over the Universe. Since the mass of $\Re[S]$ is around $m_{3/2}$ and the decay constant is small as $f_s\ll M_P$, $\Re[S]$ decays to the pNGBs and the Higgs multiplets well before the BBN.
However the decay may produce too much entropy that dilutes the preexisting baryon asymmetry. 
This moduli problem can be naturally alleviated if $\Re[S]$ also couples 
to the inflaton with operators suppressed by $1/f_s$, which is stronger than those suppressed by $1/M_P.$
In this case, the abundance of $\Re[S]$ is suppressed adiabatically~\cite{Linde:1996cx,
Nakayama:2011wqa, Nakayama:2012mf}.\footnote{There may be thermally produced scalar $\Re{[S]}$ and the fermionic partner with mass $\O(m_{3/2}),$ which decay into the MSSM particles later. This may cause the overproduction of the LSP. 
One possibility to avoid this is $f_s\lesssim 10^{16}\GEV$ so that the decay happens when the LSP is in thermal equilibrium. Also, it is avoided if the reheating temperature is small enough.  }

\section{Conclusions and discussions}

We have studied the phenomenological consequences of the supersymmetric $E_7/\SU(5)\times \U(1)^3$ non-linear sigma model. In this model, the three generations of quark and lepton chiral multiplets appear as (pseudo) Nambu Goldstone multiplets. Therefore, the origin of the three families is explained. A direct consequence is that the squarks and sleptons are massless at the tree-level. When this model couples to supergravity, a chiral multiplet $S$, which has a shift symmetry, is required. 
Hence, we have the MSSM particles and the weakly-coupled $S$ at the low energy. To break SUSY we introduce a SUSY breaking field $Z$ which is charged under some symmetry, avoiding the Polonyi problem.

We pointed out that $S$ gets a non-zero $F$-term of $f_s m_{3/2}$, which contributes to soft SUSY breaking parameters for MSSM particles. Since couplings of $S$ to Nambu Goldstone multiplets are fixed by $E_7$ invariance, the soft SUSY breaking parameters take a specific form.
In particular, the gaugino masses are dominantly generated from supersymmetry breaking mediation from $S$ multiplet and anomaly mediation at the one-loop level, and they are almost UV insensitive. 
Surprisingly, the predicted masses of the bino and wino are quasi-degenerated at $ \O(1)\%$ level 
in the region consistent with the (meta)stable electroweak vacuum.
As a result, the correct relic abundance of dark matter is explained through bino-wino coannihilation. Moreover, we found that the bottom-tau Yukawa couplings and the gauge couplings are unified precisely. The scenario can be tested at the LHC 
and can be fully covered at future collider experiments, by searching for the light gauginos as well as the light squarks and sleptons.

We have assumed that $f_s$ is around the dynamical scale of non-linear sigma and $10^{16}\,{\rm GeV} \lesssim f_s\ll M_P$. In this case, the $E_7$ symmetry and \eq{1}, may not be spoiled by non-perturbative gravity effects~\cite{Alonso:2017avz}. 
Another important issue is the SUSY CP problem. Since we have a potentially large CP violating phase in the Higgs $B$-term, on the basis that $\mu \AND m_{3/2}$ are reals, the electron electric dipole moment may exceed the experimental bound from ACME~\cite{Andreev:2018ayy}. In this paper, we have assumed that the CP violating phase is absent in the $E_7$-invariant interaction, and thus the $B$-term is treated to be real. 
A suppression mechanism of CP violating phases will be discussed elsewhere.

We have gauged the $\SU(5)$ in $H=\SU(5)\times \U(1)^3$. Alternatively, one may only gauge $\SU(3)_c\times \SU(2)_L\times \U(1)_Y$ in $\SU(5)$, which still enables us to explain the charge quantization with the partially gauged $\SU(5).$ 
In this case, the constraint on the proton decay is alleviated even for $f_s\lesssim 10^{16}\GEV$, e.g. within the QCD axion window. This is because $\SU(5)$ partners of the gauge multiplets are absent, and $\SU(5)$ partners of the Higgs doublets may not couple to the quarks and leptons. 
Then, the region with ${\rm sign}[\m]>0$, explaining the muon $g-2$ anomaly but without the bottom-tau Yukawa coupling unification, becomes more attractive. The muon $g-2$ anomaly in the context of 
NLS model will be discussed in our forthcoming work.

\section*{Acknowledgment}
We thank Keisuke Harigaya, Yutaro Shoji and Kazuya Yonekura for useful discussions.  W. Y. thanks T.D.Lee institute for the kind hospitality. 
T. T. Y. thanks Hamamatsu Photonics.
T. T. Y. is supported in part by the China Grant for Talent Scientific Start-Up Project and the JSPS Grant-in-Aid for Scientific Research No. 16H02176, and No. 17H02878, and by World Premier International Research Center Initiative (WPI Initiative), MEXT, Japan.
W.Y. is supported by NRF Strategic Research Program NRF-2017R1E1A1A01072736. N.Y. is supported by JSPS KAKENHI Grant Numbers JP15H05889, JP15K21733, and JP17H02875. 

\appendix

\section{ Supergravity potential of a NLS model}

\label{ape1}

Here, we explicitly calculate the supergravity potential of a NLS model in the K\"{a}hler manifold of $CP_1=\SU(2)/\U(1)$ to confirm our general analysis in the main part.
Let us consider the K\"{a}hler potential for the NG multiplets $\phi, \phi^\dag$ as
\begin{eqnarray}
K = F( x, y), 
\end{eqnarray}
where 
\begin{eqnarray}
x= f^2 \log{(1+\frac{\phi \phi^\dag}{f^2})}+f_s(S+S^\dag),~ y={Z^\dag Z}.
\end{eqnarray}
Here $f$ is the dynamical scale of the NLM model which does not appear in the leading order level calculation and was assumed to be the same with $f_s$ in the main part. 
We have included the SUSY breaking field $Z$, the lowest component of which is supposed to have a negligible VEV for simplicity. 
The K{\" a}hler potential for the NG multiplets is constructed from a real function transforming under $SU(2)$ as
$$
 f^2 \log{(1+\frac{\phi \phi^\dag}{f^2})}\to f^2 \log{(1+\frac{\phi \phi^\dag}{f^2})}+ f_H(\phi) + f_H(\phi)^\dag.
$$
The singlet $S$ enjoys the symmetry 
$$
 f_s S\to f_s S- f_H(\phi),
$$
so that $x$ does not change under $SU(2).$
The superpotential is given by 
\begin{eqnarray}
W= W(Z, \phi).
\end{eqnarray}
The dependence on $\phi$ implies that the $SU(2)$ is explicitly broken.

The potential from the supergravity can be derived from
\begin{eqnarray}
V(S,S^\dag, \phi, \phi^\dag, Z\simeq 0, Z^\dag\simeq 0)  &=& e^{K/M_P^2}\left|\frac{\partial K}{\partial S} \frac{W}{M_P^2} \right|^2  K^{-1}_{SS} \non \\  
&+&e^{K/M_P^2} \left(\frac{\partial K}{\partial \phi} \frac{W}{M_P^2}+\frac{\partial W}{\partial \phi}\right)\frac{\partial K}{\partial S^\dag} \frac{W}{M_P^2}  K^{-1}_{\phi S} + h.c.  \non \\
&+&e^{K/M_P^2} \left|\frac{\partial K}{\partial \phi} \frac{W}{M_P^2}+\frac{\partial W}{\partial \phi} \right|^2  K^{-1}_{\phi \phi} \label{phi} \non\\
&+&e^{K/M_P^2} \left|\frac{\partial W}{\partial Z} \right|^2  K^{-1}_{ZZ}\non \\
&-& 3  e^{K/M_P^2} \frac{|W|^2}{M_{P}^2} .
\end{eqnarray}
From straightforward calculations, one can obtain
\begin{eqnarray}
K_{\phi\phi} &=&  \frac{1}{(1+ |\phi|^2/f^2)^2} \left(K_x+ |\phi|^2 K_{xx} \right) \nonumber \\
K_{SS} &=& f_s^2 K_{xx}  \nonumber \\
K_{ZZ} &=& K_y \nonumber \\ 
K_{S\phi} &=&\frac{1}{(1+ |\phi|^2/f^2)} f_s \phi^\dag K_{xx}  \nonumber \\
K_{Z\phi} &=& 0 \nonumber \\
K_{ZS} &=&  0,
\end{eqnarray}
and 
\begin{eqnarray}
 K^{-1}_{\phi \phi} &=& (1+ |\phi|^2/f^2)^2 K_x^{-1}  \nonumber \\ 
 K^{-1}_{S S} &=&f_s^{-2} \left(|\phi|^2 K_x^{-1}+K_{xx}^{-1}\right) \nonumber \\ 
 K^{-1}_{Z Z} &=& K^{-1}_y  \nonumber \\ 
 K^{-1}_{\phi S} &=& -\phi (1+ |\phi|^2/f^2) (f_s K_x)^{-1}  \nonumber \\ 
K^{-1}_{\phi Z} &=& 0  \nonumber \\ 
 K^{-1}_{S Z} &=& 0.
\end{eqnarray}
Here ``$O'$" means $\partial O/\partial x.$ By inserting them into the potential, we arrive at
\begin{align}
\label{pottot}
V&= -e^{K/M_P^2} \frac{3|W|^2}{M_{P}^2}  \\
&+ e^{K/M_P^2} K_y^{-1} \left|\frac{\partial W}{\partial Z}\right|^2 \\
&+e^{K/M_P^2} |K_x|^2 K_{xx}^{-1}\frac{|W|^2}{M_{P}^4} \label{S} \\
&+e^{K/M_P^2}K_x^{-1} \left|\frac{\partial W}{\partial \phi}\right|^2\left(1+\frac{|\phi|^2}{f^2}\right)^2. \label{pottotla}
\end{align}
The first two terms contain the contribution of order $m_{3/2}^2 M_{P}^2$ which cancel to get a vanishingly small cosmological constant as usual. 
The third term is a potential term for the NLS model, which is of order $m_{3/2}^2 f_s^2 \ll M^2_{P} m^2_{3/2}.$
The last term represents the $F$-term contribution of $\phi$.

We emphasize that 
\begin{align}
\vev{K_x}=1, \vev{K_{xx}}= f_s^{-2}, \vev{K_y}=1
\end{align}
for $\phi$, $S, Z$ canonically normalized, respectively.

\paragraph{$A$-term}
One immediately gets that by taking $W=W_0 - y \phi^3$  
the $A$-term is generated as 
\begin{align}
A= -\frac{3 W_0^*}{M_{P}^2}.
\end{align}
Let us notice that there is no additional term canceling it. Ordinary the contribution is cancelled by the term in \eqref{phi}, but in this case this term is part of \eqref{S}, which is suppressed as \begin{align}V\supset  \frac{|W|^2}{M_P^2} \(\frac{f_s}{M_{P}}\)^2 \end{align} due to the canonical normalization of the kinetic terms.

\paragraph{Stabilization of potential by $S$}
By taking into account the kinetic normalization, let us minimize the potential by $S$ with setting $\phi,Z=0.$ 
We find the relevant terms in the potential can be expanded around the vacuum as 
\begin{align}
V\supset & e^{K/M_P^2} \left(K_y^{-1} \left|\frac{\partial W}{\partial Z}\right|^2 +|K_x|^2 K_{xx}^{-1}\frac{|W|^2}{M_{P}^4} \right) \non \\
&\simeq e^{K/M_P^2}  \left(\frac{f^2_s}{M_{P}^2} \frac{|W|^2}{M_{P}^2} +\left|\frac{\partial W}{\partial Z}\right|^2 \right)\non \\
&+  \left(2  \frac{|W|^2}{M_{P}^4}- K_{xy}\left|\frac{\partial W}{\partial Z}\right|^2-\frac{f_s^4}{M_P^4} |W|^2 K_{xxx}   \right)f_s(\delta S+\delta S^\dag) ,
\end{align}
where we have taken 
$$K_y\simeq 1+K_{xy} f_s(\delta S+\delta S^\dag),~K_x\simeq 1+f_s(\delta S+\delta S^\dag), ~K_{xx}\simeq 1+K_{xxx} f_s(\delta S+\delta S^\dag),$$
and $S=\vev{S}+\d S.$
The potential is stabilized if 
the coefficient of $\delta S+\delta S^\dag$ vanishes:
\begin{align}
\laq{stab}
 m_{3/2}^2(2-3 M_P^2 K_{xy}-f_s^4 K_{xxx})=0 ~~~  (\frac{\partial V}{\partial \Re[S]}=0).
\end{align}
Here we have used $e^{K/M_P^2} |W|^2/M_P^2 = M_P^2 m_{3/2}^2= 1/3  e^{K/M_P^2}  |\partial W /\partial Z|^2 $. 
One finds that the stabilization can be done only if $K_{xy}\neq 0$ or $K_{xxx}\neq 0$.

The mass of $\Re[S]$ can be obtained around the vacuum as 
\begin{align}
m_{\Re[S]}^2=2f_s^2\frac{\partial^2 V}{\partial x^2}= 2m_{3/2}^2+2m_{3/2}^2 f_s^2\left[ 3 M_{P}^2 (2K_{xy}^2- K_{xxy})+2f_s^2 K_{xxx} (-1+f_s^4K_{xxx})-f_s^4 K_{xxxx}
\right],\end{align}
where the higher dimensional term coefficients are evaluated at $x=\vev{x}.$

\paragraph{Summary}
In summary, the consistent non-linear sigma model should satisfy the following condition,
\begin{align}\nonumber
 (2-3 M_P^2 K_{xy}-f_s^4 K_{xxx})&=0 ~~~  (\frac{\partial V}{\partial \Re[S]}=0),\\ \nonumber
 K_x&=K_y=1,\\
\label{sum}
 K_{xx}&=f_s^{-2}.
\end{align}
\section{Formula for scalar masses}

\label{app2}
The dimension two soft parameters used in the numerical simulation at $M_{\rm inp}=10^{16}\GEV$ are given as follows for the third generation squarks and sleptons as well as the Higgs bosons.  
\begin{align}
m^2_{t_L}&= -{m_{3/2}^2\o 2} {d\o dt} \gamma_{t_L}\non \\
&-\frac{m_{3/2}}{16\pi^2 }{\sum_{a=1,2,3} C(a) g_a^2} \delta^{\rm NLS}M_a
 -\frac{m_{3/2}^2  }{8\pi^2 }\( 2+c_u\) y_t^2- \frac{m_{3/2}^2}{8\pi^2 }\( 2+c_d\) y_b^2, ~\\
m^2_{t_R}&= -{m_{3/2}^2\o 2} {d\o dt} \gamma_{t_R} \non \\ 
&-\frac{m_{3/2}}{16\pi^2 }{\sum_{a=1,2,3} C(a) g_a^2} \delta^{\rm NLS}M_a
 -2\frac{m_{3/2}^2  }{8\pi^2 }\( 2+c_u\) y_t^2,\\
\non  m^2_{b_R}&= -{m_{3/2}^2\o 2} {d\o dt} \gamma_{b_R}\\ 
&  -\frac{m_{3/2}}{16\pi^2 }\sum_{a=1,2,3} C(a) g_a^2 \delta^{\rm NLS}M_a  -2\frac{m_{3/2}^2  }{8\pi^2 }\( 2+c_d\) y_b^2,
\\\non m^2_{\t_L}&= -{m_{3/2}^2\o 2} {d\o dt} \gamma_{\tau_L}\\ &
-\frac{m_{3/2}}{16\pi^2 }\sum_{a=1,2,3} C(a) g_a^2 \delta^{\rm NLS}M_a
 -\frac{m_{3/2}^2  }{8\pi^2 }\( 2+c_d\) y_\t^2,
\\  \non m^2_{\t_R}&= -{m_{3/2}^2\o 2} {d\o dt} \gamma_{\tau_R}\\ 
&-\frac{m_{3/2}}{16\pi^2 }\sum_{a=1,2,3} C(a) g_a^2 \delta^{\rm NLS}M_a
 -2\frac{m_{3/2}^2  }{8\pi^2 }\( 2+c_d\) y_\t^2,
\\\non m^2_{{H_u}}&=-c_h m_{3/2}^2 -{m_{3/2}^2\o 2} {d\o dt} \gamma_{H_u}\\ 
&-\frac{m_{3/2}}{16\pi^2 }{\sum_{a=1,2,3} C(a) g_a^2} \delta^{\rm NLS}M_a
 -3\frac{m_{3/2}^2  }{8\pi^2 }\( 2+c_u\) y_t^2,
\\\non m^2_{{H_d}}&=-c_h m_{3/2}^2 -{m_{3/2}^2\o 2} {d\o dt} \gamma_{H_d}\\
&-\frac{m_{3/2}}{16\pi^2 }{\sum_{a=1,2,3} C(a) g_a^2} \delta^{\rm NLS}M_a  -\frac{m_{3/2}^2  }{8\pi^2 }\( 2+c_d\) (3y_d^2+y_\t^2).
\end{align}
Here 
\begin{align}
\d^{\rm NLS} M_1&= \frac{g_1^2  }{16\pi^2} m_{3/2}(12+\frac{3}{5}c_u+\frac{3}{5} c_d),\\
\d^{\rm NLS} M_2&= \frac{g_2^2  }{16\pi^2} m_{3/2}(12+c_u+ c_d),\\
\d^{\rm NLS} M_3&= \frac{g_3^2  }{16\pi^2} m_{3/2}12,
\end{align}
with $C(a)=\{Y_i^2 6/5, \d_i 3/2 ,\tl{\d}_i 8/3 \}$ for $\{\U(1)_Y, \SU(2)_L, \SU(3)_c\}$ for the supermultiplet $i$. 
 $\d_i=1$ ($\tl{\d}_i=1$) for an $\SU(2)_L$ doublet ($\SU(3)_c$ triplet) otherwise $0$.  $\g_i$ can be found in Ref.~\cite{Martin:1993zk}.
 For a first two generation squark or slepton, we use the formula for the third generation by taking the Yukawa couplings to zero.


\begin{thebibliography}{99}

\bibitem{Kugo:1983ai} 
  T.~Kugo and T.~Yanagida,
  Phys.\ Lett.\  {\bf 134B}, 313 (1984).
  doi:10.1016/0370-2693(84)90007-8


\bibitem{Yanagida:1985jc} 
  T.~Yanagida and Y.~Yasui,
  Nucl.\ Phys.\ B {\bf 269}, 575 (1986).
  doi:10.1016/0550-3213(86)90512-2


\bibitem{Irie:1983cd} 
  S.~Irie and Y.~Yasui,
  Z.\ Phys.\ C {\bf 29}, 123 (1985).
  doi:10.1007/BF01571392


\bibitem{Ibe:2006de} 
  M.~Ibe, T.~Moroi and T.~T.~Yanagida,
  Phys.\ Lett.\ B {\bf 644}, 355 (2007)
  doi:10.1016/j.physletb.2006.11.061
  [hep-ph/0610277].


\bibitem{Ibe:2011aa} 
  M.~Ibe and T.~T.~Yanagida,
  Phys.\ Lett.\ B {\bf 709}, 374 (2012)
  doi:10.1016/j.physletb.2012.02.034
  [arXiv:1112.2462 [hep-ph]].


\bibitem{ArkaniHamed:2012gw} 
  N.~Arkani-Hamed, A.~Gupta, D.~E.~Kaplan, N.~Weiner and T.~Zorawski,
  arXiv:1212.6971 [hep-ph].


\bibitem{Yanagida:2016kag} 
  T.~T.~Yanagida, W.~Yin and N.~Yokozaki,
  JHEP {\bf 1609}, 086 (2016)
  doi:10.1007/JHEP09(2016)086
  [arXiv:1608.06618 [hep-ph]].


\bibitem{Yamaguchi:2016oqz} 
  M.~Yamaguchi and W.~Yin,
  PTEP {\bf 2018}, no. 2, 023B06 (2018)
  doi:10.1093/ptep/pty002
  [arXiv:1606.04953 [hep-ph]].


\bibitem{Giudice:1998xp} 
  G.~F.~Giudice, M.~A.~Luty, H.~Murayama and R.~Rattazzi,
  JHEP {\bf 9812}, 027 (1998)
  doi:10.1088/1126-6708/1998/12/027
  [hep-ph/9810442].


\bibitem{Randall:1998uk} 
  L.~Randall and R.~Sundrum,
  Nucl.\ Phys.\ B {\bf 557}, 79 (1999)
  doi:10.1016/S0550-3213(99)00359-4
  [hep-th/9810155].


\bibitem{Yin:2016shg} 
  W.~Yin and N.~Yokozaki,
  Phys.\ Lett.\ B {\bf 762}, 72 (2016)
  doi:10.1016/j.physletb.2016.09.024
  [arXiv:1607.05705 [hep-ph]].


\bibitem{Yanagida:2018eho} 
  T.~T.~Yanagida, W.~Yin and N.~Yokozaki,
  JHEP {\bf 1804}, 012 (2018)
  doi:10.1007/JHEP04(2018)012
  [arXiv:1801.05785 [hep-ph]].


\bibitem{Komargodski:2010rb} 
  Z.~Komargodski and N.~Seiberg,
  JHEP {\bf 1007}, 017 (2010)
  doi:10.1007/JHEP07(2010)017
  [arXiv:1002.2228 [hep-th]].


\bibitem{Kugo:2010fs} 
  T.~Kugo and T.~T.~Yanagida,
  Prog.\ Theor.\ Phys.\  {\bf 124}, 555 (2010)
  doi:10.1143/PTP.124.555
  [arXiv:1003.5985 [hep-th]].


\bibitem{Goto:1990me} 
  T.~Goto and T.~Yanagida,
  Prog.\ Theor.\ Phys.\  {\bf 83}, 1076 (1990).
  doi:10.1143/PTP.83.1076


\bibitem{Siegel:1978mj} 
  W.~Siegel and S.~J.~Gates, Jr.,
  Nucl.\ Phys.\ B {\bf 147}, 77 (1979).
  doi:10.1016/0550-3213(79)90416-4


\bibitem{Kugo:1982cu} 
  T.~Kugo and S.~Uehara,
  Nucl.\ Phys.\ B {\bf 226}, 49 (1983).
  doi:10.1016/0550-3213(83)90463-7


\bibitem{Kugo:1983mv} 
  T.~Kugo and S.~Uehara,
  Prog.\ Theor.\ Phys.\  {\bf 73}, 235 (1985).
  doi:10.1143/PTP.73.235


\bibitem{Sato:1997hv} 
  J.~Sato and T.~Yanagida,
  Phys.\ Lett.\ B {\bf 430}, 127 (1998)
  doi:10.1016/S0370-2693(98)00510-3
  [hep-ph/9710516].


\bibitem{Yanagida:1979as} 
  T.~Yanagida,
  Conf.\ Proc.\ C {\bf 7902131}, 95 (1979).


\bibitem{GellMann:1980vs} 
  M.~Gell-Mann, P.~Ramond and R.~Slansky,
  Conf.\ Proc.\ C {\bf 790927}, 315 (1979)
  [arXiv:1306.4669 [hep-th]].


\bibitem{Glashow:1979nm} 
  S.~L.~Glashow,
  NATO Sci.\ Ser.\ B {\bf 61}, 687 (1980).
  doi:10.1007/978-1-4684-7197-7\_15


\bibitem{Minkowski:1977sc} 
  P.~Minkowski,
  Phys.\ Lett.\  {\bf 67B}, 421 (1977).
  doi:10.1016/0370-2693(77)90435-X


\bibitem{Inoue:1991rk} 
  K.~Inoue, M.~Kawasaki, M.~Yamaguchi and T.~Yanagida,
  Phys.\ Rev.\ D {\bf 45}, 328 (1992).
  doi:10.1103/PhysRevD.45.328


\bibitem{Bagger:1999rd} 
  J.~A.~Bagger, T.~Moroi and E.~Poppitz,
  JHEP {\bf 0004}, 009 (2000)
  doi:10.1088/1126-6708/2000/04/009
  [hep-th/9911029].


\bibitem{Boyda:2001nh} 
  E.~Boyda, H.~Murayama and A.~Pierce,
  Phys.\ Rev.\ D {\bf 65}, 085028 (2002)
  doi:10.1103/PhysRevD.65.085028
  [hep-ph/0107255].


\bibitem{Choi:2004sx} 
  K.~Choi, A.~Falkowski, H.~P.~Nilles, M.~Olechowski and S.~Pokorski,
  JHEP {\bf 0411}, 076 (2004)
  doi:10.1088/1126-6708/2004/11/076
  [hep-th/0411066].


\bibitem{Choi:2005ge} 
  K.~Choi, A.~Falkowski, H.~P.~Nilles and M.~Olechowski,
  Nucl.\ Phys.\ B {\bf 718}, 113 (2005)
  doi:10.1016/j.nuclphysb.2005.04.032
  [hep-th/0503216].


\bibitem{Endo:2005uy} 
  M.~Endo, M.~Yamaguchi and K.~Yoshioka,
  Phys.\ Rev.\ D {\bf 72}, 015004 (2005)
  doi:10.1103/PhysRevD.72.015004
  [hep-ph/0504036].


\bibitem{Chowdhury:2015rja} 
  D.~Chowdhury and N.~Yokozaki,
  JHEP {\bf 1508}, 111 (2015)
  doi:10.1007/JHEP08(2015)111
  [arXiv:1505.05153 [hep-ph]].


\bibitem{Binetruy:2000md} 
  P.~Binetruy, M.~K.~Gaillard and B.~D.~Nelson,
  Nucl.\ Phys.\ B {\bf 604}, 32 (2001)
  doi:10.1016/S0550-3213(00)00759-8
  [hep-ph/0011081].


\bibitem{Evans:2013uza} 
  J.~L.~Evans, M.~Ibe, K.~A.~Olive and T.~T.~Yanagida,
  Eur.\ Phys.\ J.\ C {\bf 74}, no. 2, 2775 (2014)
  doi:10.1140/epjc/s10052-014-2775-9
  [arXiv:1312.1984 [hep-ph]].


\bibitem{Riotto:1995am} 
  A.~Riotto and E.~Roulet,
  Phys.\ Lett.\ B {\bf 377}, 60 (1996)
  doi:10.1016/0370-2693(96)00313-9
  [hep-ph/9512401].


\bibitem{Djouadi:2002ze} 
  A.~Djouadi, J.~L.~Kneur and G.~Moultaka,
  Comput.\ Phys.\ Commun.\  {\bf 176}, 426 (2007)
  doi:10.1016/j.cpc.2006.11.009
  [hep-ph/0211331].


\bibitem{Kusenko:1996jn} 
  A.~Kusenko, P.~Langacker and G.~Segre,
  Phys.\ Rev.\ D {\bf 54}, 5824 (1996)
  doi:10.1103/PhysRevD.54.5824
  [hep-ph/9602414].


\bibitem{Pierce:1996zz} 
  D.~M.~Pierce, J.~A.~Bagger, K.~T.~Matchev and R.~j.~Zhang,
  Nucl.\ Phys.\ B {\bf 491}, 3 (1997)
  doi:10.1016/S0550-3213(96)00683-9
  [hep-ph/9606211].


\bibitem{Vega:2015fna} 
  J.~Pardo Vega and G.~Villadoro,
  JHEP {\bf 1507}, 159 (2015)
  doi:10.1007/JHEP07(2015)159
  [arXiv:1504.05200 [hep-ph]].


\bibitem{Draper:2013oza} 
  P.~Draper, G.~Lee and C.~E.~M.~Wagner,
  Phys.\ Rev.\ D {\bf 89}, no. 5, 055023 (2014)
  doi:10.1103/PhysRevD.89.055023
  [arXiv:1312.5743 [hep-ph]].


\bibitem{Baer:2005jq} 
  H.~Baer, T.~Krupovnickas, A.~Mustafayev, E.~K.~Park, S.~Profumo and X.~Tata,
  JHEP {\bf 0512}, 011 (2005)
  doi:10.1088/1126-6708/2005/12/011
  [hep-ph/0511034].


\bibitem{ArkaniHamed:2006mb} 
  N.~Arkani-Hamed, A.~Delgado and G.~F.~Giudice,
  Nucl.\ Phys.\ B {\bf 741}, 108 (2006)
  doi:10.1016/j.nuclphysb.2006.02.010
  [hep-ph/0601041].


\bibitem{Ibe:2013pua} 
  M.~Ibe, A.~Kamada and S.~Matsumoto,
  Phys.\ Rev.\ D {\bf 89}, no. 12, 123506 (2014)
  doi:10.1103/PhysRevD.89.123506
  [arXiv:1311.2162 [hep-ph]].


\bibitem{Harigaya:2014dwa} 
  K.~Harigaya, K.~Kaneta and S.~Matsumoto,
  Phys.\ Rev.\ D {\bf 89}, no. 11, 115021 (2014)
  doi:10.1103/PhysRevD.89.115021
  [arXiv:1403.0715 [hep-ph]].


\bibitem{Duan:2018rls} 
  G.~H.~Duan, K.~I.~Hikasa, J.~Ren, L.~Wu and J.~M.~Yang,
  Phys.\ Rev.\ D {\bf 98}, no. 1, 015010 (2018)
  doi:10.1103/PhysRevD.98.015010
  [arXiv:1804.05238 [hep-ph]].


\bibitem{Cohen:2013xda} 
  T.~Cohen, T.~Golling, M.~Hance, A.~Henrichs, K.~Howe, J.~Loyal, S.~Padhi and J.~G.~Wacker,
  JHEP {\bf 1404}, 117 (2014)
  doi:10.1007/JHEP04(2014)117
  [arXiv:1311.6480 [hep-ph]].


\bibitem{Abe:2019egv} 
  T.~Abe, S.~Chigusa, Y.~Ema and T.~Moroi,
  Phys.\ Rev.\ D {\bf 100}, no. 5, 055018 (2019)
  doi:10.1103/PhysRevD.100.055018
  [arXiv:1904.11162 [hep-ph]].


\bibitem{Arvanitaki:2014wva} 
  A.~Arvanitaki, M.~Baryakhtar and X.~Huang,
  Phys.\ Rev.\ D {\bf 91}, no. 8, 084011 (2015)
  doi:10.1103/PhysRevD.91.084011
  [arXiv:1411.2263 [hep-ph]].


\bibitem{Cardoso:2018tly} 
  V.~Cardoso, \'{O}.~J.~C.~Dias, G.~S.~Hartnett, M.~Middleton, P.~Pani and J.~E.~Santos,
  JCAP {\bf 1803}, 043 (2018)
  doi:10.1088/1475-7516/2018/03/043
  [arXiv:1801.01420 [gr-qc]].


\bibitem{Stott:2018opm} 
  M.~J.~Stott and D.~J.~E.~Marsh,
  Phys.\ Rev.\ D {\bf 98}, no. 8, 083006 (2018)
  doi:10.1103/PhysRevD.98.083006
  [arXiv:1805.02016 [hep-ph]].


\bibitem{Graham:2018jyp} 
  P.~W.~Graham and A.~Scherlis,
  Phys.\ Rev.\ D {\bf 98}, no. 3, 035017 (2018)
  doi:10.1103/PhysRevD.98.035017
  [arXiv:1805.07362 [hep-ph]].


\bibitem{Guth:2018hsa} 
  F.~Takahashi, W.~Yin and A.~H.~Guth,
  Phys.\ Rev.\ D {\bf 98}, no. 1, 015042 (2018)
  doi:10.1103/PhysRevD.98.015042
  [arXiv:1805.08763 [hep-ph]].


\bibitem{Ho:2019ayl} 
  S.~Y.~Ho, F.~Takahashi and W.~Yin,
  JHEP {\bf 1904}, 149 (2019)
  doi:10.1007/JHEP04(2019)149
  [arXiv:1901.01240 [hep-ph]].


\bibitem{Daido:2017wwb} 
  R.~Daido, F.~Takahashi and W.~Yin,
  JCAP {\bf 1705}, 044 (2017)
  doi:10.1088/1475-7516/2017/05/044
  [arXiv:1702.03284 [hep-ph]].


\bibitem{Daido:2017tbr} 
  R.~Daido, F.~Takahashi and W.~Yin,
  JHEP {\bf 1802}, 104 (2018)
  doi:10.1007/JHEP02(2018)104
  [arXiv:1710.11107 [hep-ph]].


\bibitem{Takahashi:2019qmh} 
  F.~Takahashi and W.~Yin,
  JHEP {\bf 1907}, 095 (2019)
  doi:10.1007/JHEP07(2019)095
  [arXiv:1903.00462 [hep-ph]].


\bibitem{Ouellet:2019tlz} 
  J.~L.~Ouellet {\it et al.},
  Phys.\ Rev.\ D {\bf 99}, no. 5, 052012 (2019)
  doi:10.1103/PhysRevD.99.052012
  [arXiv:1901.10652 [physics.ins-det]].


\bibitem{Silva-Feaver:2016qhh} 
  M.~Silva-Feaver {\it et al.},
  IEEE Trans.\ Appl.\ Supercond.\  {\bf 27}, no. 4, 1400204 (2017)
  doi:10.1109/TASC.2016.2631425
  [arXiv:1610.09344 [astro-ph.IM]].


\bibitem{Irastorza:2018dyq} 
  I.~G.~Irastorza and J.~Redondo,
  Prog.\ Part.\ Nucl.\ Phys.\  {\bf 102}, 89 (2018)
  doi:10.1016/j.ppnp.2018.05.003
  [arXiv:1801.08127 [hep-ph]].


\bibitem{Salemi:2019xgl} 
  C.~P.~Salemi [ABRACADABRA Collaboration],
  arXiv:1905.06882 [hep-ex].


\bibitem{Linde:1996cx} 
  A.~D.~Linde,
  Phys.\ Rev.\ D {\bf 53}, R4129 (1996)
  doi:10.1103/PhysRevD.53.R4129
  [hep-th/9601083].


\bibitem{Nakayama:2011wqa} 
  K.~Nakayama, F.~Takahashi and T.~T.~Yanagida,
  Phys.\ Rev.\ D {\bf 84}, 123523 (2011)
  doi:10.1103/PhysRevD.84.123523
  [arXiv:1109.2073 [hep-ph]].


\bibitem{Nakayama:2012mf} 
  K.~Nakayama, F.~Takahashi and T.~T.~Yanagida,
  Phys.\ Lett.\ B {\bf 714}, 256 (2012)
  doi:10.1016/j.physletb.2012.06.072
  [arXiv:1203.2085 [hep-ph]].


\bibitem{Alonso:2017avz} 
  R.~Alonso and A.~Urbano,
  JHEP {\bf 1902}, 136 (2019)
  doi:10.1007/JHEP02(2019)136
  [arXiv:1706.07415 [hep-ph]].


\bibitem{Andreev:2018ayy} 
  V.~Andreev {\it et al.} [ACME Collaboration],
  Nature {\bf 562}, no. 7727, 355 (2018).
  doi:10.1038/s41586-018-0599-8


\bibitem{Martin:1993zk} 
  S.~P.~Martin and M.~T.~Vaughn,
  Phys.\ Rev.\ D {\bf 50}, 2282 (1994)
  Erratum: [Phys.\ Rev.\ D {\bf 78}, 039903 (2008)]
  doi:10.1103/PhysRevD.50.2282, 10.1103/PhysRevD.78.039903
  [hep-ph/9311340].
  
    \end{thebibliography}
\end{document}